\documentclass[aps,showpacs,preprint,superscriptaddress]{revtex4}
\usepackage{graphicx}
\usepackage{subfig}
\usepackage{amsmath}
\usepackage{amssymb}
\usepackage{ulem}
\usepackage{bm}
\usepackage{float}
\begin{document}
\title{Massless quark production in chromoelectric field with back reaction}
\author{M. R. Jia}
\author{F. Wan}
\author{C. Lv}
\affiliation{Key Laboratory of Beam Technology and Materials Modification of the Ministry of Education, and College of Nuclear Science and Technology, Beijing Normal University, Beijing 100875, China}
\author{B. S. Xie \footnote{Corresponding author. Email address: bsxie@bnu.edu.cn}}
\affiliation{Key Laboratory of Beam Technology and Materials Modification of the Ministry of Education, and College of Nuclear Science and Technology, Beijing Normal University, Beijing 100875, China}
\affiliation{Beijing Radiation Center, Beijing 100875, China}

\date{\today}%

\begin{abstract}
We study the massless quark production in $SU(2)$ gauge chromoelectric field by single-time Wigner function covariantly with back reaction. The evolution of field and current are investigated. For a phenomenological distribution function, both the time and momentum dependence have been studied. Interesting phenomena are found, which are: when considering the back reaction, the yield of quark production is higher than the Bjorken expanding field, and  momentum 'gap' with confinement phenomenon exists in the phenomenological distribution function. To have a better understanding on the phenomena, components of the Wigner function are qualitatively analyzed.
\end{abstract}
\pacs{24.85.+p, 25.75.-q, 12.38.Mh}

\maketitle

\section{Introduction}
During the past years, a large amount of experiments have been carried out in relativistic heavy ion collisions at the Super-Proton Synchrotron(SPS) and at the Relativistic Heavy Ion Collier(RHIC). These experiments are thought to produce quark-gluon plasmas, and related theories have been investigated\cite{Heinz1,Heinz2,Heinz3} in early years. Since the microscopic mechanisms of hadron production in hadron-hadron and heavy ion collisions are not fully understood, it is very important to improve our understandings on these problems.

According to the chromoelectric flux tube("string") models, see Refs.\cite{Andersson,Wang1,Pop,Sorge,Marco1,Marco2,Heiben,Mace}, hadrons will be produced via quark-antiquark and diquark-antidiquark pair production from the field energy, namely, from the unstable flux tubes. These models can describe experimental results very successfully at small transverse momentum. At higher transverse momentum one can apply perturbative QCD models\cite{Field,Wang2,Zhang}. Since the energies of the string density is expected to be large enough at RHIC and LHC, that a strong collective gluon field will be formed in the whole available transverse volume. Thus, a classical gluon field as the expectation value of the quantum field can be considered and investigated in the problem. The properties of such nonAbelian classical fields and details of gluon production were studied very intensively, especially asymptotic solutions\cite{McLerran,Gelis}. Lattice calculations were performed to describe strong classical fields under finite space-time conditions in the very early stage of heavy ion collision\cite{Krasnitz,Lappi,Gelis1}. Similar methods have been developed to investigate the influence of inhomogeneity on particle production\cite{Gies2,Kim,Dunne}. Both fermion and boson production were calculated in different kinetic models, see Refs.\cite{Gatoff1,Gatoff2,Kluger,Wong,Eisenberg,Vinnik,Skokov1,Skokov2,Skokov3,Dietrich,Nayak}. Some comparison have been made with $U(1)$ case in the kinetic model, see Refs.\cite{Skokov4,Skokov5}. These calculations concentrated mostly on the bulk properties of the gluon and quark matter, the time evolution, the time dependence of energy and particle number densities and so on.

To improve our understanding on the process of early quark production, we calculate the kinetic equation of Wigner Function with the back reaction in $SU(2)$ gauge field. Our main interest is focused on the difference between a system with back reaction and one without. Here, we focus our attention on a fixed color direction for massless fermions. To calculate the problem, it comes to face the full $p$ region problem, naturally. It is hard to deal with both the nonperturbative and perturbative models  at the same time. In this work, we just simplify the problem and focus on finding some properties in both nonperturbative and perturbative models. We hope the model will meet some specific physical condition. Though, it seems finding a proper function $g(p_{\perp},p_{3})$ (function of the 'coupling constant') may be a possible way to deal with this problem, this may come in our future consideration.

The paper is orgnized as follows. In Sec. \uppercase\expandafter{\romannumeral2}, we show the QCD-QFT problem by introducing the kinetic equation of the Wigner operator and the field equation. In Sec. \uppercase\expandafter{\romannumeral3}, we give the formulas of the Wigner function under the assumptions mentioned in Refs.\cite{Skokov3,Elze,Heinz4,Skokov4}. In Sec. \uppercase\expandafter{\romannumeral4}, we simplify the self-consistent system in a $SU(2)$ gauge, where the 4-potential has been fixed in a special direction. The current, the boundary condition and the phenomenological distribution function are also introduced in this section. In Sec. \uppercase\expandafter{\romannumeral5}, we give the numerical results and show the similarities and new phenomena comparing with nonself-consistent system. In Sec. \uppercase\expandafter{\romannumeral6}, we summarize our results and make some outlooks.
\section{Kinetic Equation of  Wigner Operator}
\noindent
The colored quarks obey the Dirac equation.
\begin{equation}
\{\gamma^{\mu}\left[i\partial_{\mu}+gA_{\mu}(x)\right]-m\}\psi(x)=0.
\end{equation}
Here $m$ denotes the current mass of quarks, $g$ is the coupling constant. After choosing the metric $g^{\mu\nu}=(1,-1,-1,-1)$, and choosing unit such that $\hbar=c=1$, the gauged field potential is an $N\times N$ matrix in color space defined by
\begin{equation}
A_{\mu}(x)=A_{\mu}^{a}(x)t^{a}\label{eq1},
\end{equation}
with the $N^{2}-1$ hermitian generators ($t^{a}$) of $SU(N)$ in the fundamental representation, satisfying $Tr(t^{a})=0, Tr(t^{a}t^{b})=\frac{1}{2}\delta^{ab}$, and $\left[t^{a},t^{b}\right]=if^{abc}t^{c}$, with $f^{abc}$ is the structure constant, the covariant derivative reads
\begin{equation}
\mathcal{D}_{\mu}(x)\equiv\partial_{\mu}+igA_{\mu}(x)\label{eq2},
\end{equation}
is an $N\times N$ matrix in color space, field strength tensor is $F_{\mu\nu}(x)\equiv\left[\mathcal{D}_{\mu}(x),\mathcal{D}_{\nu}(x)\right]/(ig)$, which obeys the field equation
\begin{equation}
\left[\mathcal{D}_{\mu}(x),F^{\mu\nu}(x)\right]=g\mathcal{J}^{\nu}(x)\label{eq3}.
\end{equation}
where $\mathcal{J}^{\mu}\equiv\hat{j}^{\mu,a}t^{a}\equiv \bar{\psi}\gamma^{\mu}t^{a}\psi t^{a}=\int d^{4}pTr(\gamma^{\mu}t^{a}\hat{\mathcal{W}}(x,p))t^{a}$, with $\hat{\mathcal{W}}(x,p)$ is the Wigner operator.
The covariant Wigner operator is defined as
\begin{eqnarray}
\begin{aligned}
\hat{\mathcal{W}}(x,p)
&\equiv\int \frac{d^4y}{(2\pi)^4}e^{-ip\cdot y}\bar{\psi}(x+\frac{1}{2}y)\mathcal{U}(x+\frac{1}{2}y,x)\otimes\mathcal{U}(x,x-\frac{1}{2}y)\psi(x-\frac{1}{2}y)\\
&=\int \frac{d^4y}{(2\pi)^2}e^{-ip\cdot y}\bar{\psi}(x)e^{y\cdot \frac{1}{2}\partial_{x}^{\dag}}\otimes e^{-y\cdot\frac{1}{2}\partial_{x}}\psi(x)\label{eq5},
\end{aligned}
\end{eqnarray}
with $\partial_{x}^{\dag}\equiv\partial/\overleftarrow{\partial}x$ and $\partial_{x}\equiv\partial/\overrightarrow{\partial}x$ are the generators of translations acting to the left and right respectively. The link operator $\mathcal{U}(\beta,\alpha)$ is given by the path ordered exponential of a line integral\cite{Itzykson,Becher},
\begin{equation}
\mathcal{U}(\beta,\alpha)\equiv\mathcal{P}\mathrm{exp}\left(-ig\int\limits_{\alpha}^{\beta}dz^{\mu}A_{\mu}^{a}(z)t^{a}\right)\label{eq4}
\end{equation}
and the path of integration is chosen as straight line between the end points, $z(s)\equiv z(\beta,\alpha,s)=\alpha+(\beta-\alpha)s$, with $0\leqslant s\leqslant 1$.
The Schwinger string is defined as \cite{,Heinz4}
\begin{eqnarray}
^{[x]}\hat{F}_{\mu\nu}(z(s))\equiv \mathcal{U}(x,z(s))\hat{F}_{\mu\nu}(z(s))\mathcal{U}(z(s),x)=\mathrm{exp}\left[is\partial_{p}^{\alpha}\widetilde{\mathcal{D}}_{\alpha}(x)\right]\hat{F}_{\mu\nu}(x)\label{eq6},
\end{eqnarray}
where $\widetilde{\mathcal{D}}(x)\hat{F}_{\mu\nu}(x)\equiv\left[\mathcal{D}(x),\hat{F}_{\mu\nu}(x)\right]$, and then we obtain the covariant transport equation for the QCD quark Wigner operator reads
\begin{equation}
\begin{aligned}
 \left[\gamma^{\mu}p_{\mu}-m+\frac{1}{2}i\gamma^{\mu}\mathcal{D}_{\mu}(x)\right]
 & \hat{\mathcal{W}}(x,p)=\\
 -ig\partial^{\nu}_{p}\gamma^{\mu}
 &\{\int\limits_{0}^{\frac{1}{2}}ds(1-2s)^{[x]}F_{\mu\nu}(x+is\partial_{p})\hat{\mathcal{W}}(x,p)+\\
 &\hat{\mathcal{W}}(x,p)\int\limits_{-\frac{1}{2}}^{0}ds(1-2s)^{[x]}F_{\mu\nu}(x+is\partial_{p})\}.
\end{aligned}
\label{eq7}
\end{equation}
Now Eq.(\ref{eq3}) and Eq.(\ref{eq7}) constitute a self-consistent system, which is another form of the QCD-QFT problem with back reaction.

\section{Kinetic Equation of Wigner Function}
In order to simplify this problem, we choose a special field, which is only in the $z$ direction and depends only on $t$. Then the gauged 4-potential reads
\begin{equation}
A_{\mu}^{a}(t)=(0,0,0,A^{a}(t))\label{eq8}.
\end{equation}

We also assume that the Wigner function is sufficiently smooth in momentum space and the field strength varies slowly enough in coordinate space, which satisfies $(\varDelta p)_{\mathcal{W}}\cdot(\varDelta x)_{F}\gg \hbar$ \cite{Heinz4}. Moreover, the expectation value of the Wigner operator is assumed to be diagonal in the gauge that diagonalizes the field tensor\cite{Elze}. Then, make  the first order approximation for slow varying field, a space homogeneous Wigner function, which is investigated by the kinetic equation in the frame of the covariant single-time formalism, and can be read\cite{Skokov4}
\begin{equation}
\begin{aligned}
\partial_{t}\mathcal{W}
&+\frac{g}{8}\frac{\partial}{\partial p_{i}}(4\{\mathcal{W},F_{0i}\}
+2\{F_{i\nu},\left[\mathcal{W},\gamma^{0}\gamma^{\nu}\right]\}\\
&-\left[F_{i\nu},\{\mathcal{W},\gamma^{0}\gamma^{\nu}\}\right])=ip_{i}\{\gamma^{0}\gamma^{i},\mathcal{W}\}-im\left[\gamma^{0},\mathcal{W}\right]\\
&+ig\left[A_{i},\left[\gamma^{0}\gamma^{i},\mathcal{W}\right]\right].
\end{aligned}
\label{eq9}	
\end{equation}
The color decomposition with $SU(N_{c})$ generators in fundamental representation is
\begin{equation}
\mathcal{W}= \mathcal{W}^{s}+\mathcal{W}^{a}t^{a},\quad a=1,2,....,N_{c}^{2}-1\label{eq10},
\end{equation}
where $\mathcal{W}^{s}$ is the singlet part and $\mathcal{W}^{a}$ is the multiplet part. Also the spinor decomposition follows
\begin{equation}
\mathcal{W}^{s|a}=a^{s|a}+\mathbf{b}_{\mu}^{s|a}\gamma^{\mu}+\mathbf{c}_{\mu\nu}^{s|a}\sigma ^{\mu\nu}+\mathbf{d}_{\mu}^{s|a}\gamma^{\mu}\gamma^{5}+ie^{s|a}\gamma^{5}\label{eq11}.
\end{equation}
In the frame of the specific 4-potential Eq.(\ref{eq8}), Eq.(\ref{eq3}) can be reduced to
\begin{equation}
\partial_{t}E^{a}(t)=-\mathcal{J}^{a}(t)\label{eq12},
\end{equation}
where after defining
\begin{equation}
\partial_{t}A^{a}(t)=-E^{a}(t)\label{eq13},
\end{equation}
then the problem comes to a Maxwell-like one.
\section{Kinetic Equations For Wigner Function in $SU(2)$ case}
\subsection{Kinetic Equations For $SU(2)$ Case}
For $SU(2)$ case, substituting both color decompositions Eq.(\ref{eq10}) and spinor decomposition Eq.(\ref{eq11}) into Eq.(\ref{eq9}). One can obtain a system of coupled differential equations, which consists of 32 components\cite{Skokov4}. For massless fermions, where $m\approx0$. In a fixed color direction field, we obtain
\begin{eqnarray}
&& \partial_t\mathbf{b}^{s}+\frac{3}{4}gE^{a}\frac{\partial}{\partial p_{3}}\mathbf{b}^{a}=2\mathbf{p}\times\mathbf{d}^{s},\label{eq20}\\
&& \partial_t\mathbf{b}^{a}+gE^{a}\frac{\partial}{\partial p_{3}}\mathbf{b}^{s}=2\mathbf{p}\times\mathbf{d}^{a},\label{eq21}\\
&& \partial_t\mathbf{d}^{s}+\frac{3}{4}gE^{a}\frac{\partial}{\partial p_{3}}\mathbf{d}^{a}=2\mathbf{p}\times\mathbf{b}^{s},\label{eq22}\\
&& \partial_t\mathbf{d}^{a}+gE^{a}\frac{\partial}{\partial p_{3}}\mathbf{d}^{s}=2\mathbf{p}\times\mathbf{b}^{a},\label{eq23}
\end{eqnarray}
call a vector decomposition in momentum space,
\begin{eqnarray}
&& \mathbf{b}^{s|a}=b^{s|a}_{3}\mathbf{n}+b^{s|a}_{\perp}\frac{\mathbf{p}_{\perp}}{p_{\perp}},\\
&& \mathbf{d}^{s|a}=d^{s|a}\mathbf{n}\times\frac{\mathbf{p}_{\perp}}{p_{\perp}},
\end{eqnarray}
Eq.(\ref{eq20})-(\ref{eq23}) can be reduced to 6 equations, and read
\begin{eqnarray}
&&\partial_{t}b_{\perp}^{s}+\frac{3g}{4}E^{a}\frac{\partial}{\partial p_{3}}b_{\perp}^{a}=-2p_{3}d^{s},\label{eq14}\\
&&\partial_{t} b_{3}^{s}+\frac{3g}{4}E^{a}\frac{\partial}{\partial p_{3}}b_{3}^{a}=2p_{\perp}d^{s},\label{eq15}\\
&&\partial_{t}d^{s}+\frac{3g}{4}E^{a}\frac{\partial}{\partial p_{3}}d^{a}=2p_{3}b_{\perp}^{s}-2p_{\perp}b_{3}^{s},\label{eq16}\\
&&\partial_{t}b_{\perp}^{a}+gE^{a}\frac{\partial}{\partial b_{\perp}^{s}}=-2p_{3}d^{a},\label{eq17}\\
&&\partial_{t}b_{3}^{a}+gE^{a}\frac{\partial}{\partial p_{3}}b_{3}^{s}=2p_{\perp}d^{a},\label{eq18}\\
&&\partial_{t}d^{a}+gE^{a}\frac{\partial}{\partial p_{3}}d^{s}=2p_{3}b_{\perp}^{a}-2p_{\perp}b_{3}^{a}\label{eq19}.
\end{eqnarray}
\subsection{The Quark Current}
To calculate Eq.(\ref{eq14})-(\ref{eq19}) with considering the back reaction, the field Eq.(\ref{eq12}) and Eq.(\ref{eq13}) should be taken account of. And $\mathcal{J}(t)$ can be written in components
\begin{eqnarray}
&& j_{\nu}^{s}(t)=-Tr(\gamma_{\nu}\mathbf{1}\bar{\psi}(t)\psi(t))=\int d^{4}p Tr(\gamma_{\nu}\mathbf{1}\mathcal{W}(t,\mathbf{p})),\label{eq24}\\
&& j_{\nu}^{a}(t)=-Tr(\gamma_{\nu}t_{a}\bar{\psi}(t)\psi(t))=\int d^{4}pTr(\gamma_{\nu}t^{a}\mathcal{W}(t,\mathbf{p}))\label{eq25}.
\end{eqnarray}
Since we only consider the case that the field is only in a fixed color direction, the singlet component of the current does not affect. Substituting the decomposition Eq.(\ref{eq10}) and Eq.(\ref{eq11}) with Dirac Matrix and Pauli Matrix, after integration, we have only the vector component $\mathbf{b}_{\mu}$ remained, while other components $a, \mathbf{c}_{\mu\nu}, \mathbf{d}_{\mu}, e$ vanish.

\subsection{The Vaccum Solutions of the Wigner Fucntion and The Phenomenological Distribution Function}
The vacuum solution for the singlet Wigner function has the form\cite{Skokov3}
\begin{equation}
\mathcal{W}^{s}=-\frac{1}{2}\frac{m+\mathbf{p}\cdot\gamma}{\omega(\mathbf{p})},
\end{equation}
here, $\omega(\mathbf{p})=\sqrt{\mathbf{p}^{2}+m^{2}}$, the vacuum initial conditions can be given as
\begin{equation}
\mathbf{b}^{s}(t\rightarrow-\infty)=-\frac{\mathbf{p}}{2|\mathbf{p}|},
\end{equation}
while other components of the Wigner function have zero initial values.

Comparing the one-particle energy density from Wigner Function $\mathcal{E}_{f}(t)=Tr\langle (m-\gamma^{i}p_{i})\mathcal{W}(t,\mathbf{p})+\omega(\mathbf{p})\rangle$ and from distribution function $\mathcal{E}_{f}(t)=4N_{c}\int\frac{d^3\mathbf{p}}{(2\pi)^3}\omega(\mathbf{p})f_{f}(t,\mathbf{p})$, see Refs.\cite{Skokov3,Skokov4,Heinz4}, one can obtain a phenomenological distribution function for fermions $f_{f}(t,\mathbf{p})$, reads
\begin{equation}
f_{f}(t,\mathbf{p})=\frac{ma^{s}(t,\mathbf{p})+\mathbf{p}\cdot\mathbf{b}^{s}(t,\mathbf{p})}{\omega(\mathbf{p})}+\frac{1}{2}\label{eq30},
\end{equation}
which is positive defined in a nonzero field.
Moreover, $f_{f}=0$ agrees with the vacuum solution of the Wigner function. Which is physically correct.
\section{Numerical Results}
In this section, we solve Eq.(\ref{eq14})-(\ref{eq19}) with Eq.(\ref{eq12})-(\ref{eq13}) numerically. As we mentioned in Sec. \uppercase\expandafter{\romannumeral1} and Sec. \uppercase\expandafter{\romannumeral4}, to obtain the current $\mathcal{J}$, we need to integrate Eq.(\ref{eq24})-(\ref{eq25}) in whole $p$ region. Then, a proper cut off should be taken. Also, the amplitude of the field $E_{0}$ and the coupling constant $g$ affect the convergent region badly. So, the field amplitude and the coupling constant are specially chosen, and we set them as $E_{0}=0.34$ and $g=1$. To calculate the problem, we set $E(t\leqslant0)=E_{0}\cdot\left[1-\mathrm{tanh}^{2}(t/\delta)\right]$, with $\delta=0.1E_{0}^{1/2}/|E_{0}|$. And when $t>0$, the system evolute in a self-consistent way. To make a comparison, the Bjorken expanding field is introduced, which has a same opening at $t\leqslant0$, and $E(t>0)=\frac{E_{0}}{(1+t/t_{0})^{\kappa}}$ with $t_{0}=0.01E_{0}^{1/2}/|E_{0}|$ and $\kappa=2/3$. We list the components of the boundary conditions as
\begin{eqnarray}
&&\mathbf{b}^{s}_{3}(t\rightarrow-\infty)=-\frac{p_{3}}{2|\mathbf{p}|},\\
&&\mathbf{b}^{s}_{\perp}(t\rightarrow-\infty)=-\frac{p_{\perp}}{2|\mathbf{p}|},\\
&&\mathbf{b}^{a}_{3}(t\rightarrow-\infty)=\mathbf{b}^{a}_{\perp}(t\rightarrow-\infty)=0,\\
&&\mathbf{d}^{a}_{3}(t\rightarrow-\infty)=\mathbf{d}^{a}_{\perp}(t\rightarrow-\infty)=0.
\end{eqnarray}

\begin{figure}[H]
  \centering
  \subfloat[Field]{
  \includegraphics[width=0.9\linewidth]{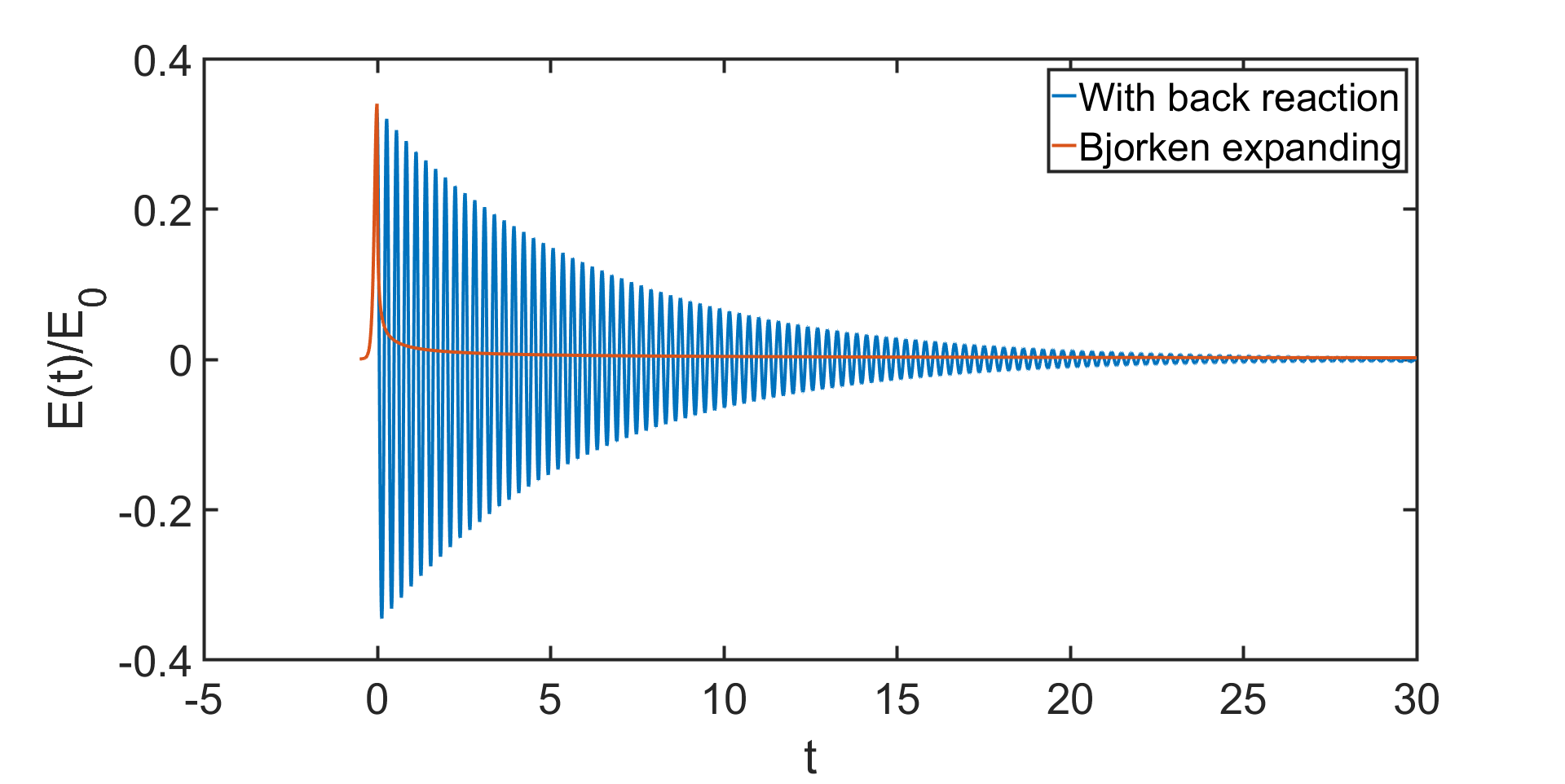}}\\
  \subfloat[Current]{
  \includegraphics[width=0.9\linewidth]{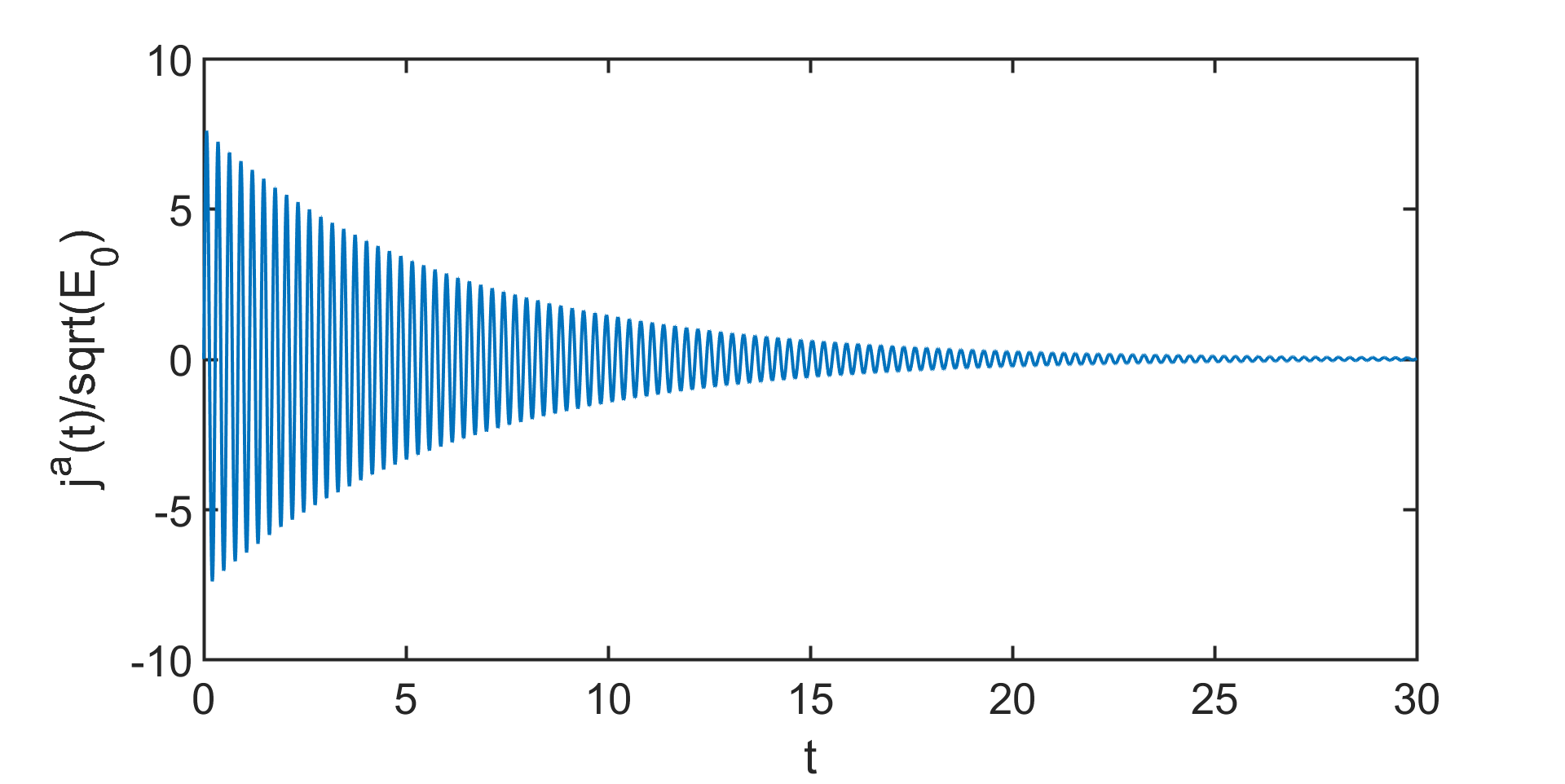}}
  \caption{Evolution of the Field and Current, $t$ is in unit of $\frac{|E_{0}|}{\sqrt{E_{0}}}$}
  \label{fig1}
\end{figure}

In Fig.\ref{fig1}, we plot evolution of the field and the current. When considering the back reaction, the current is a sensitive quantity, which is characteristic plotted in Fig.\ref{fig1}(b). Both the field and current are damping, and the back reaction field does not decrease that fast than the Bjorken expanding field, which indicates that the relaxation time can not be ignored. Thus, the typical time of the quark production will be different.

\begin{figure}[H]
  \centering
  \includegraphics[width=0.9\linewidth]{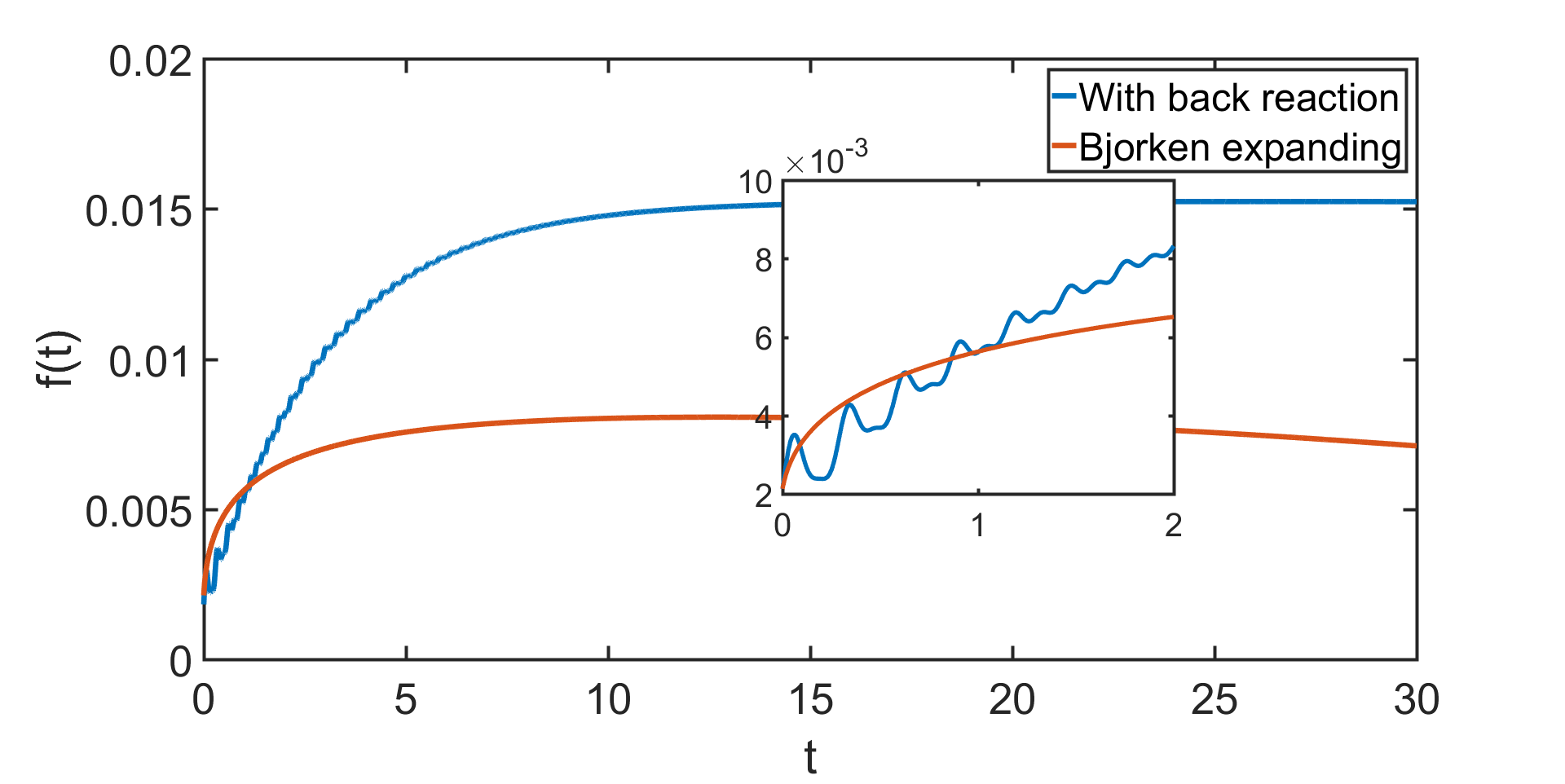}
  \caption{Evolution of the phenomenological distribution function $f(t)$, $t$ is in unit of $\frac{|E_{0}|}{\sqrt{E_{0}}}$}
  \label{fig2}
\end{figure}

To study the time dependence of the quark production, we plot the evolution of the phenomenological distribution function in Fig.\ref{fig2}. Comparing distribution function in back reaction field ($f_{br}(t)$) and in Bjorken expanding field ($f_{Bjorken}(t)$), it is found that, when $t<1$, $f_{Bjorken}(t)>f_{br}(t)$, but when $t>1$, $f_{Bjorken}(t)<f_{br}(t)$, and when $t$ is large enough, they both reach a equilibrium state(while the little decrease in Bjorken expanding field is cause by the computational accuracy). In a classical view, this agrees with the result in Fig.\ref{fig1}, when considering the back reaction, yield of massless quark is higher because of the field decreasing more slowly.

\begin{figure}[H]
  \centering
  \subfloat[t=0.1]{
  \includegraphics[width=0.45\linewidth]{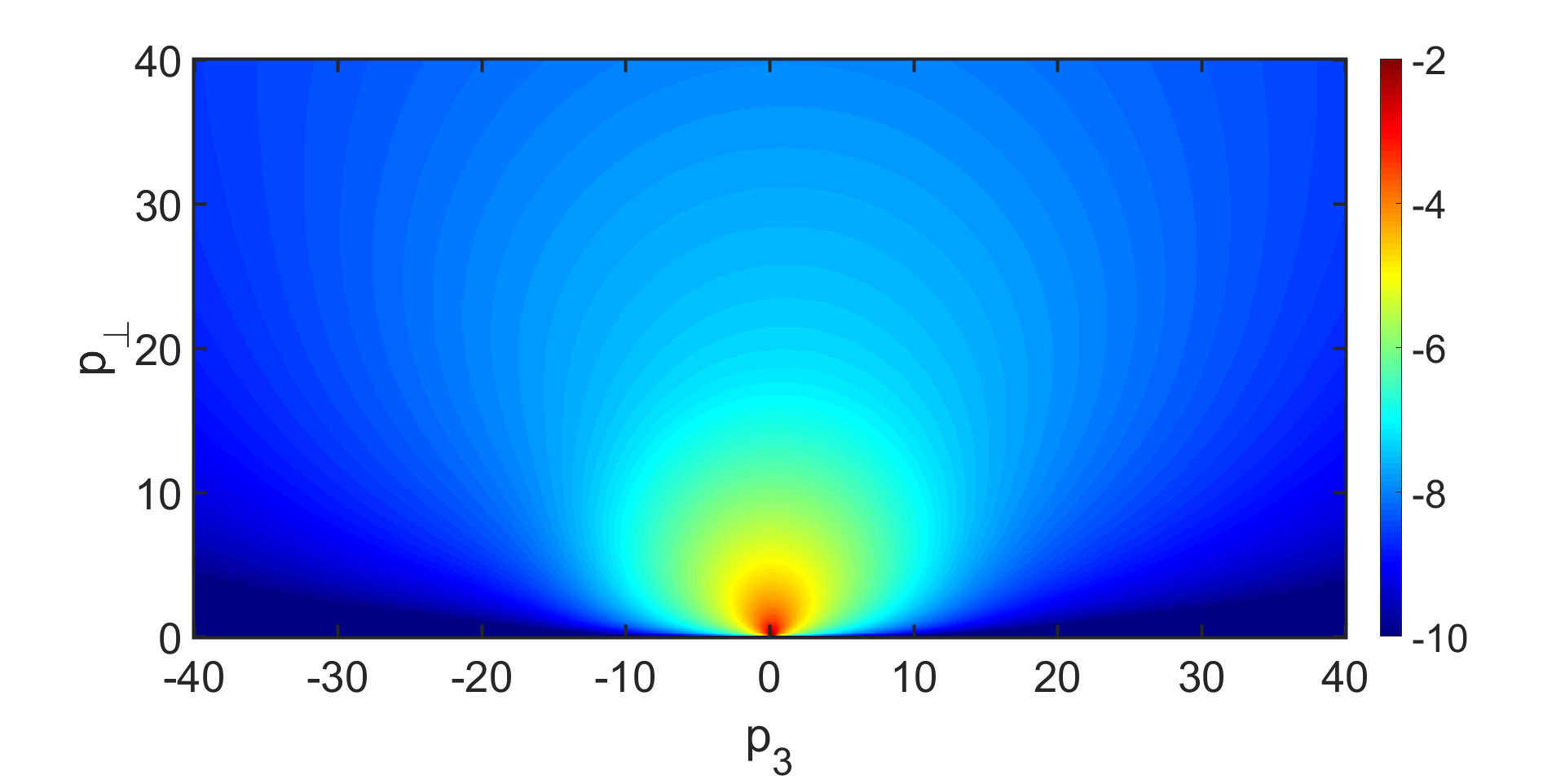}}\
  \subfloat[t=0.5]{
  \includegraphics[width=0.45\linewidth]{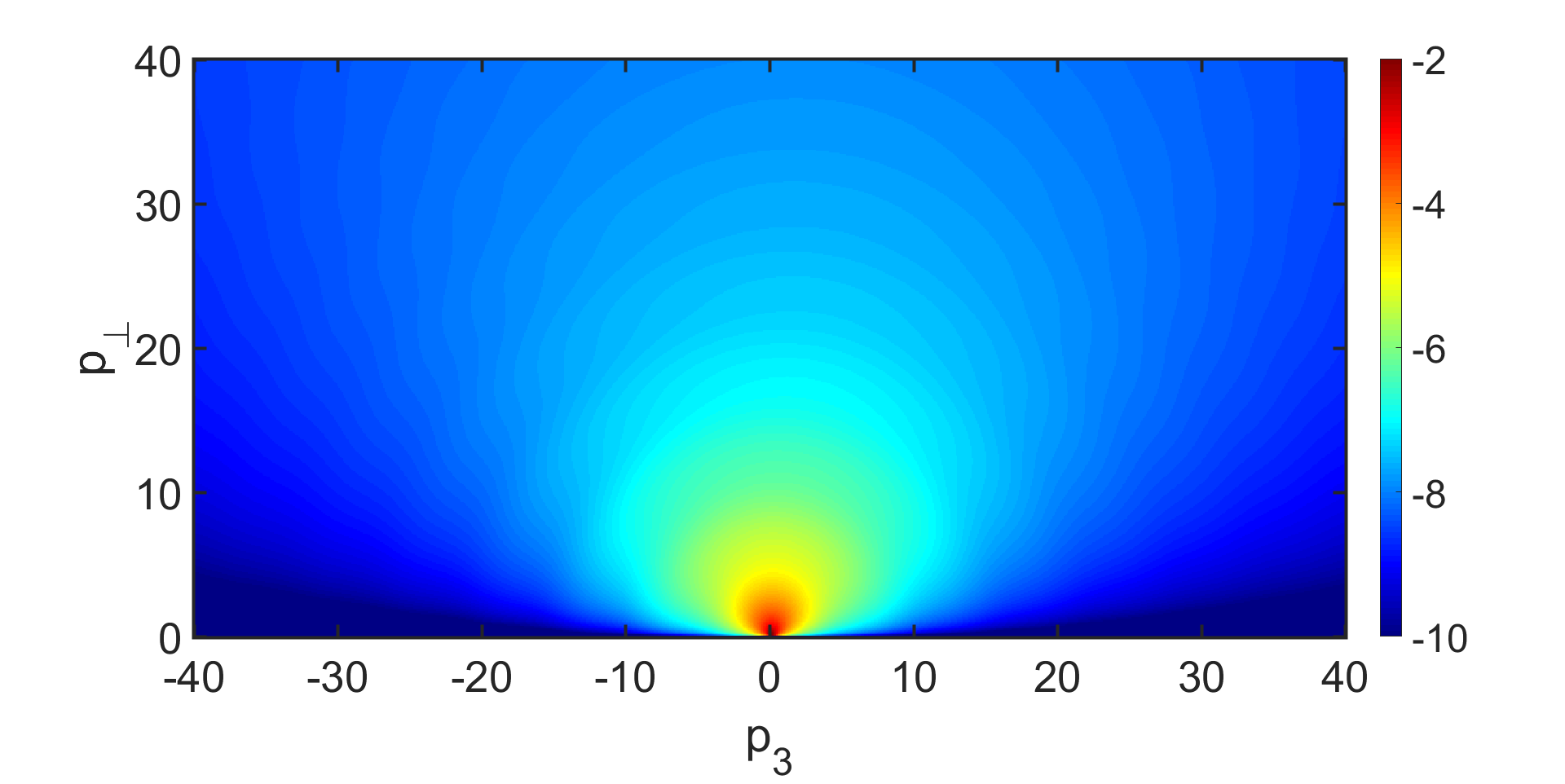}}\\
  \subfloat[t=1.0]{
  \includegraphics[width=0.45\linewidth]{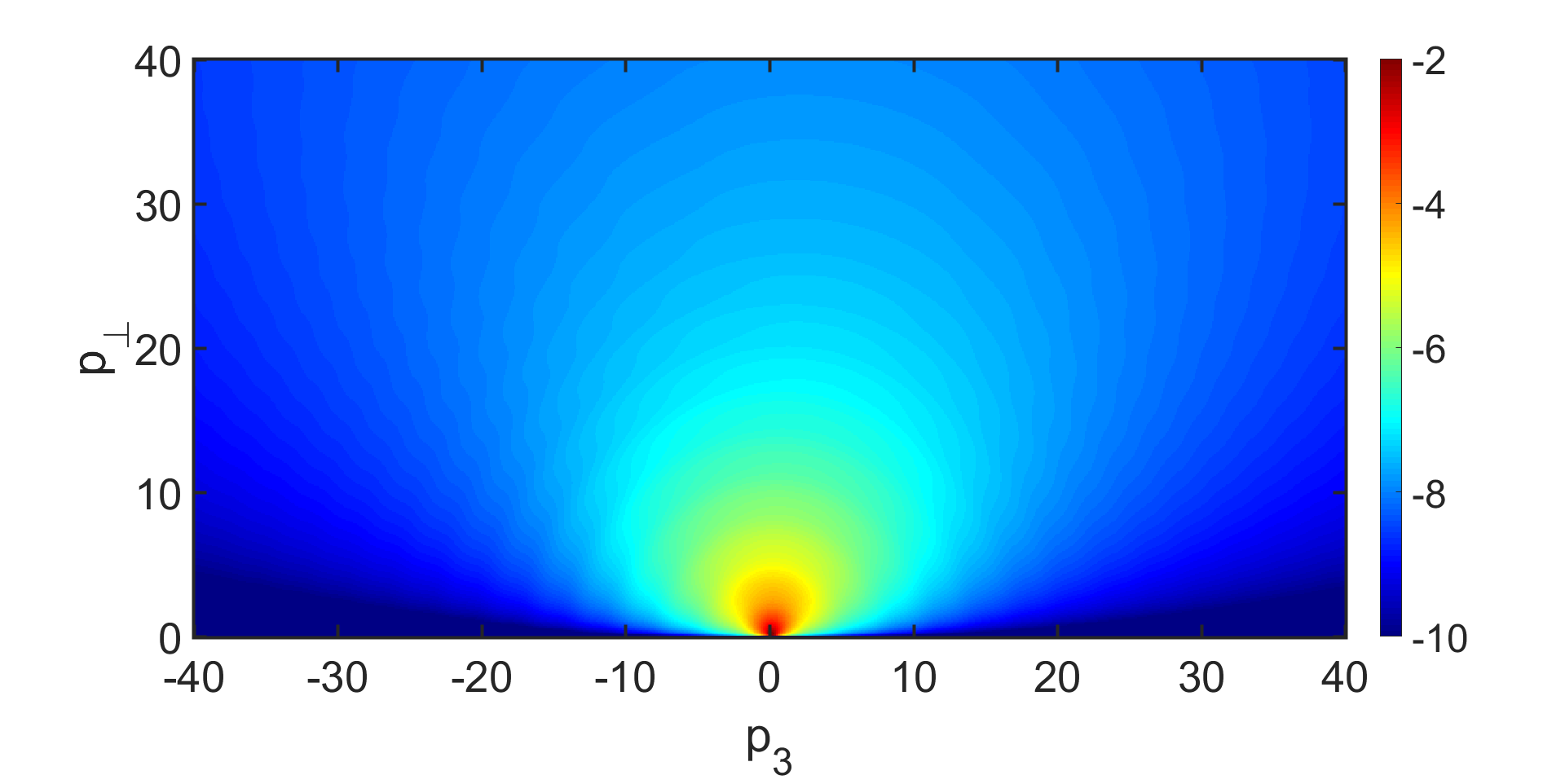}}\
  \subfloat[t=2.0]{
  \includegraphics[width=0.45\linewidth]{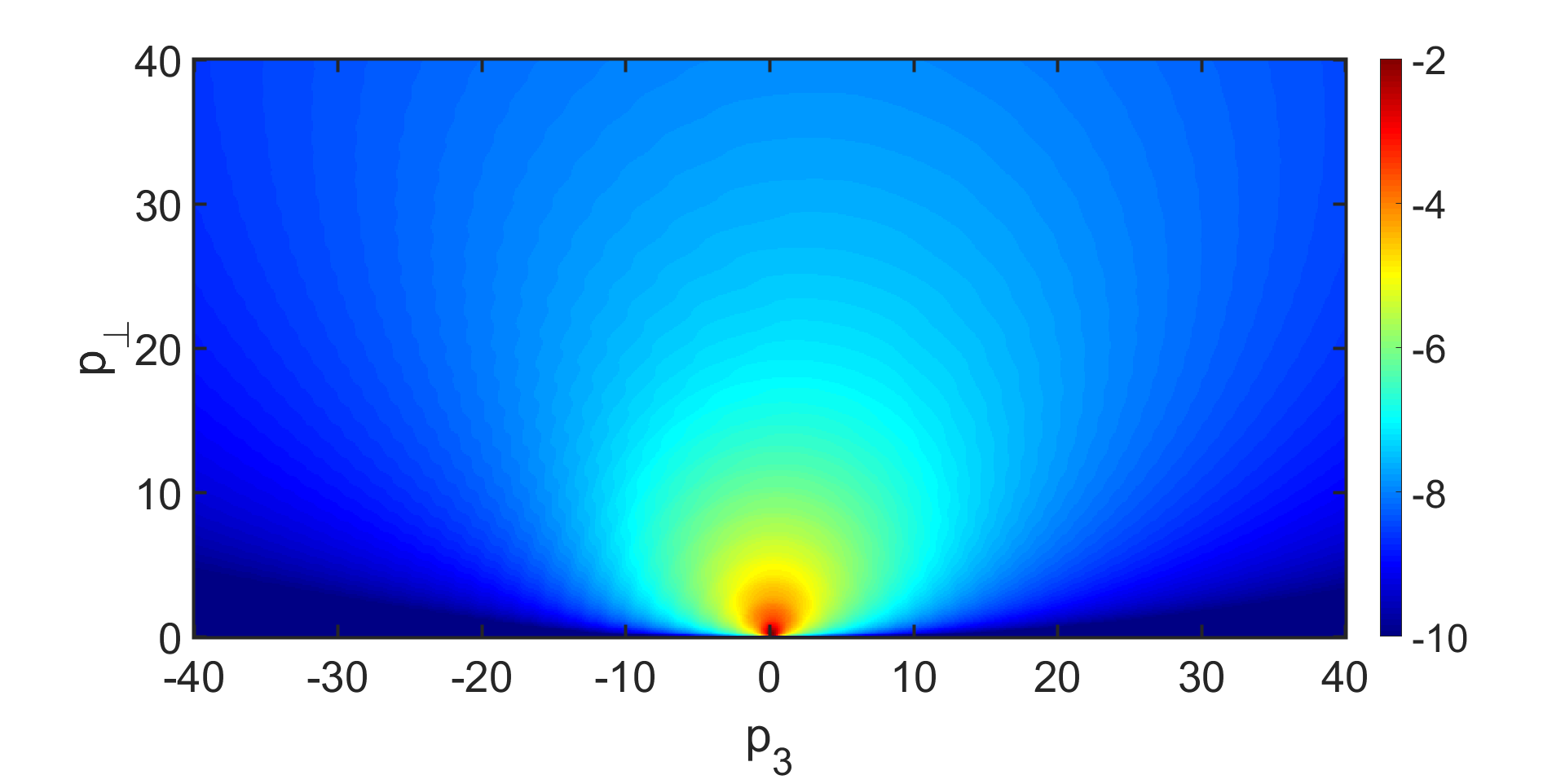}}
  \caption{(color on line)The phenomenological distribution function $Log(f_{Bjorken}(t))$ in the Bjorken expanding field at different time points, where both $p_{\perp}$ and $p_{3}$ are in unit of $\frac{|E_{0}|}{\sqrt{E_{0}}}$}
  \label{fig3}
\end{figure}

\begin{figure}[H]
  \centering
  \subfloat[t=0.1]{
  \includegraphics[width=0.45\linewidth]{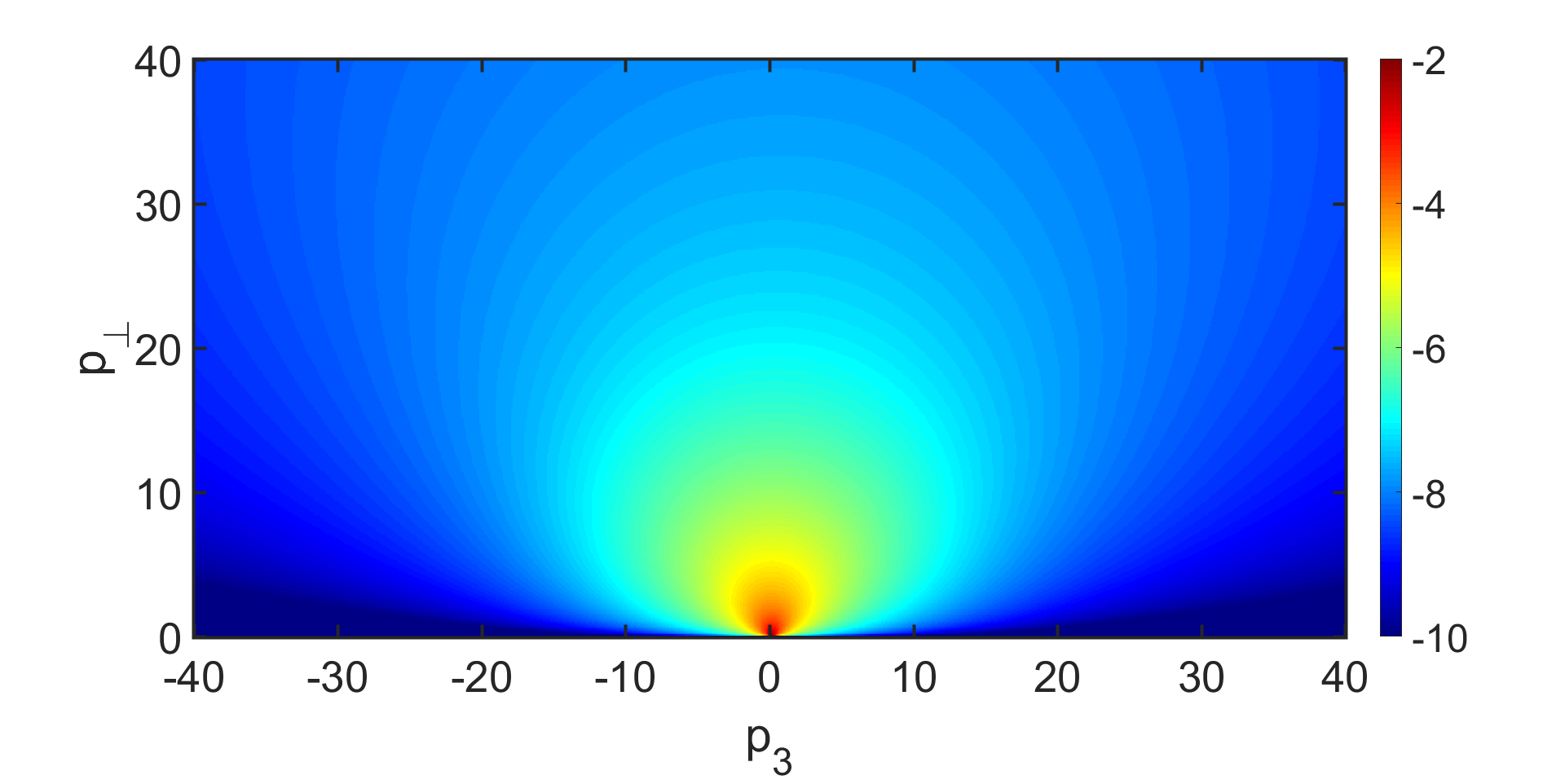}}\
  \subfloat[t=0.3]{
  \includegraphics[width=0.45\linewidth]{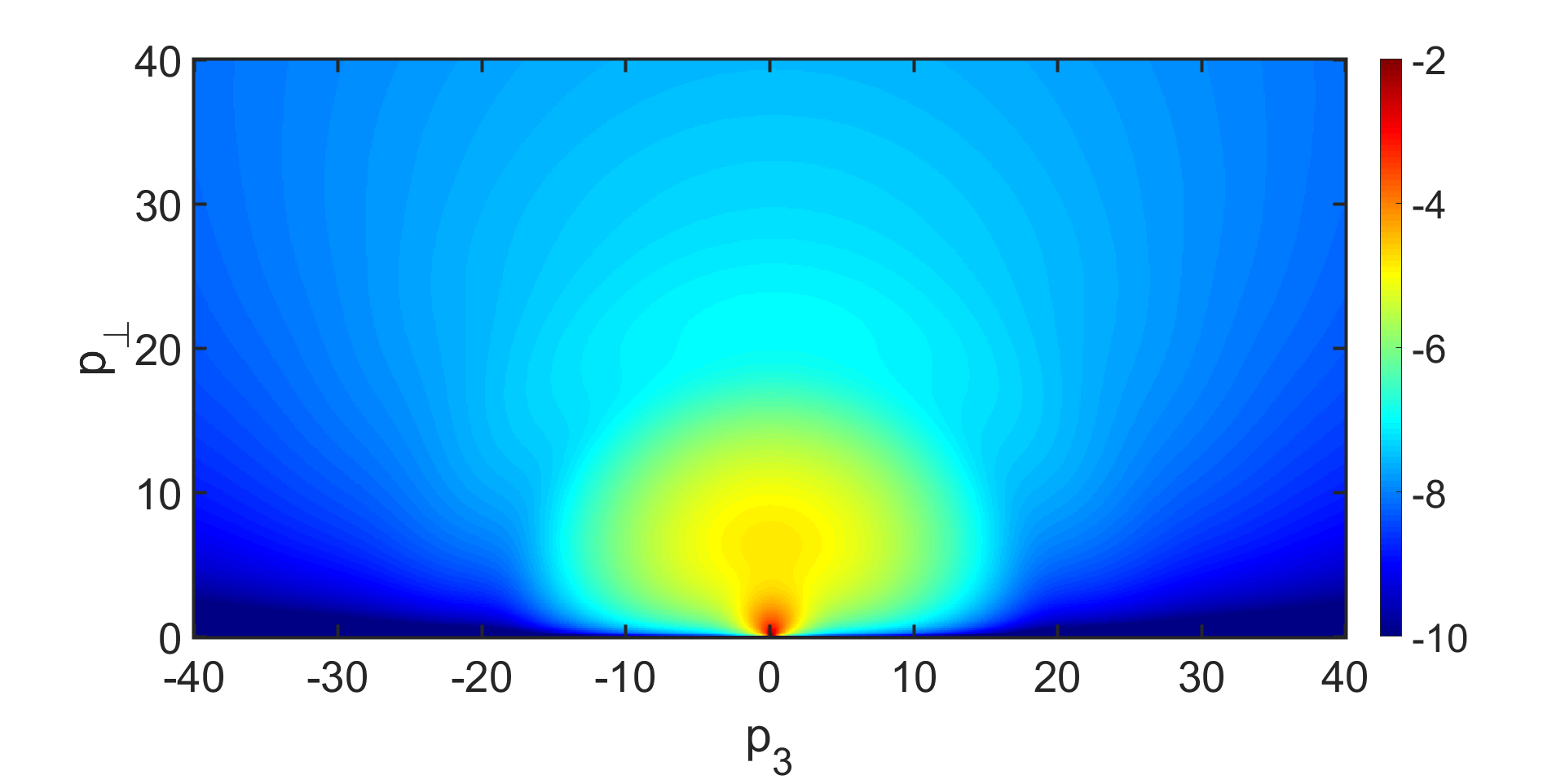}}\\
  \subfloat[t=0.5]{
  \includegraphics[width=0.45\linewidth]{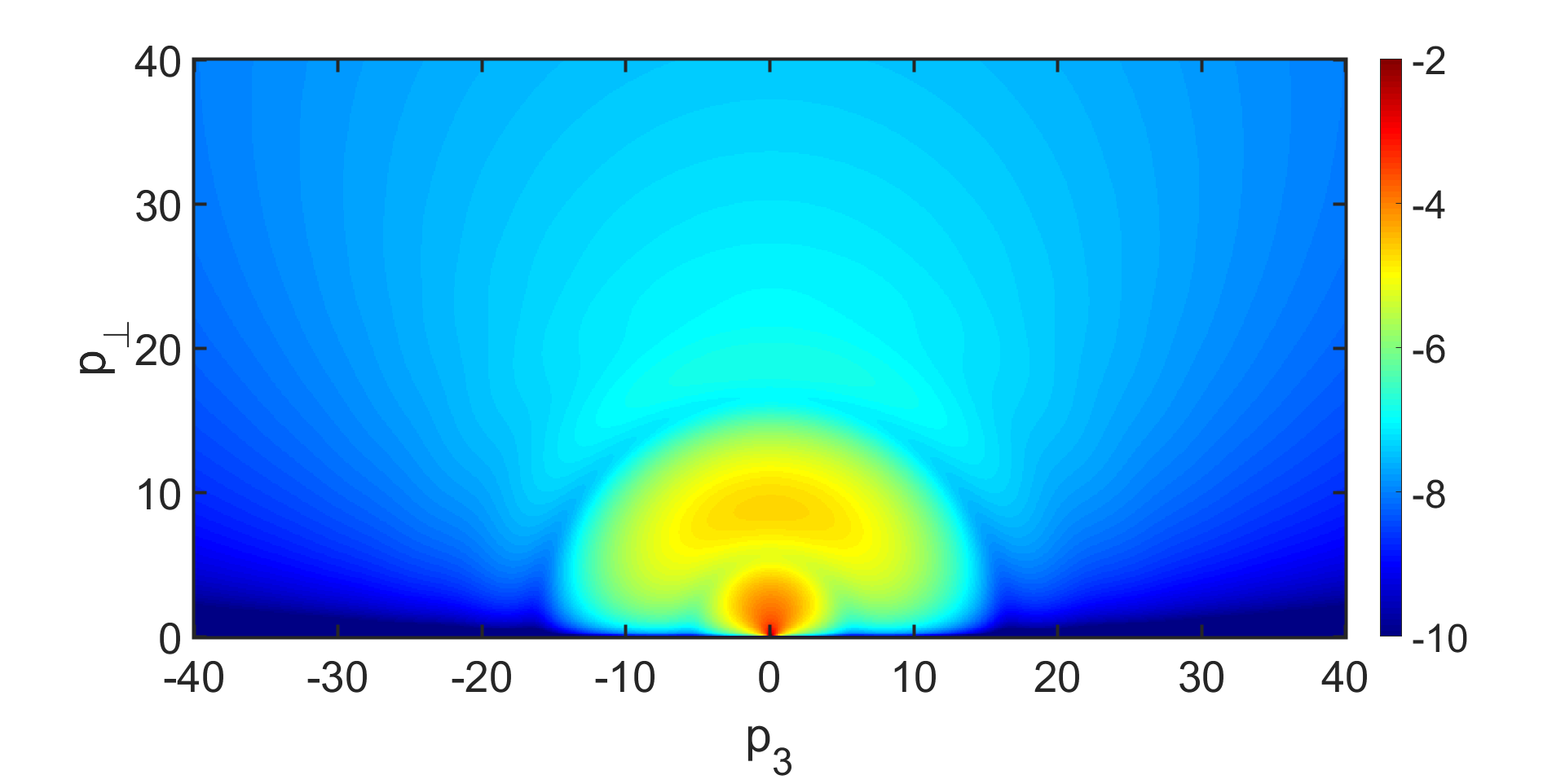}}\
  \subfloat[t=0.7]{
  \includegraphics[width=0.45\linewidth]{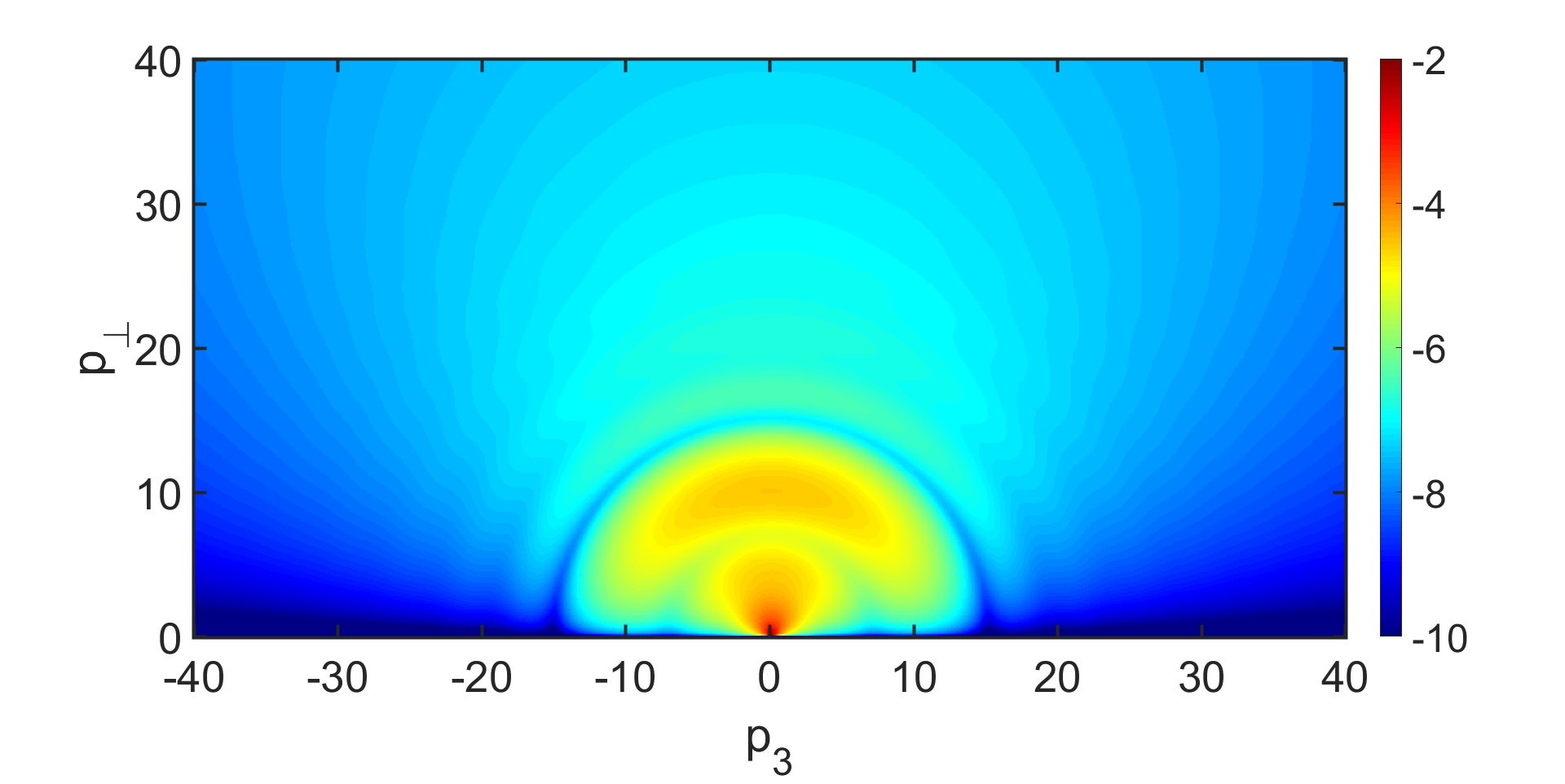}}\\
  \subfloat[t=0.9]{
  \includegraphics[width=0.45\linewidth]{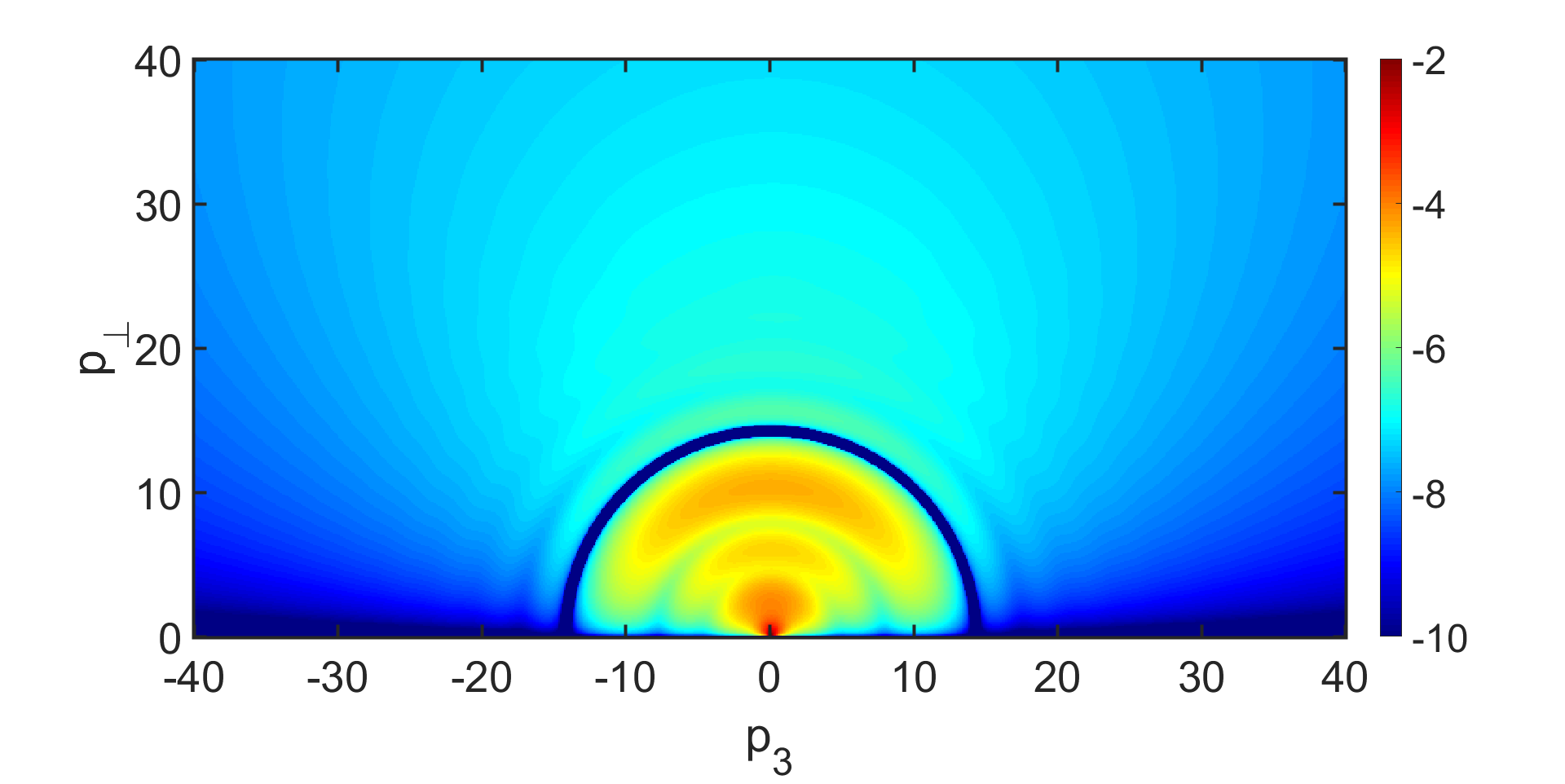}}\
  \subfloat[t=2.0]{
  \includegraphics[width=0.45\linewidth]{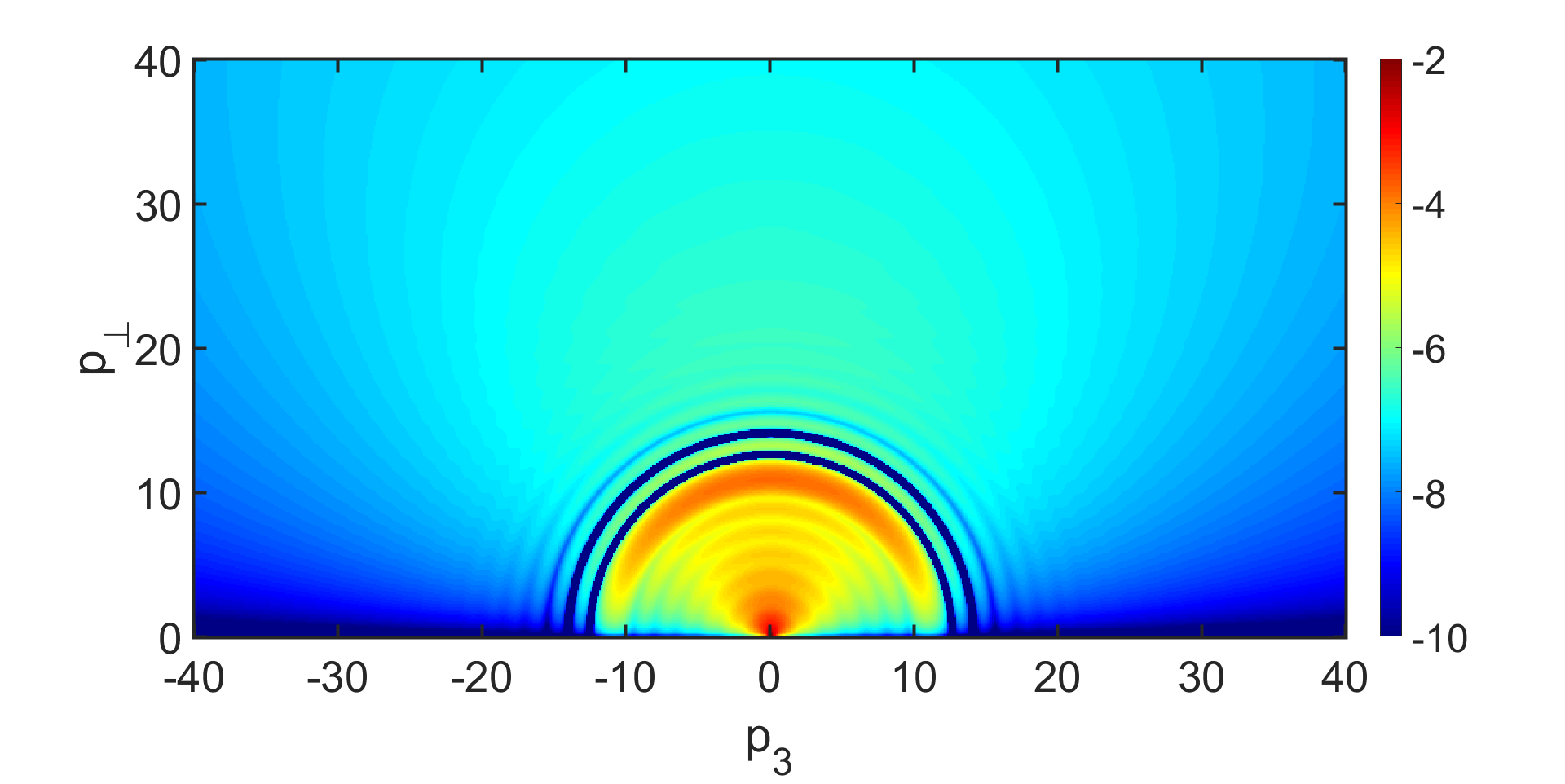}}\\
  \subfloat[t=5.0]{
  \includegraphics[width=0.45\linewidth]{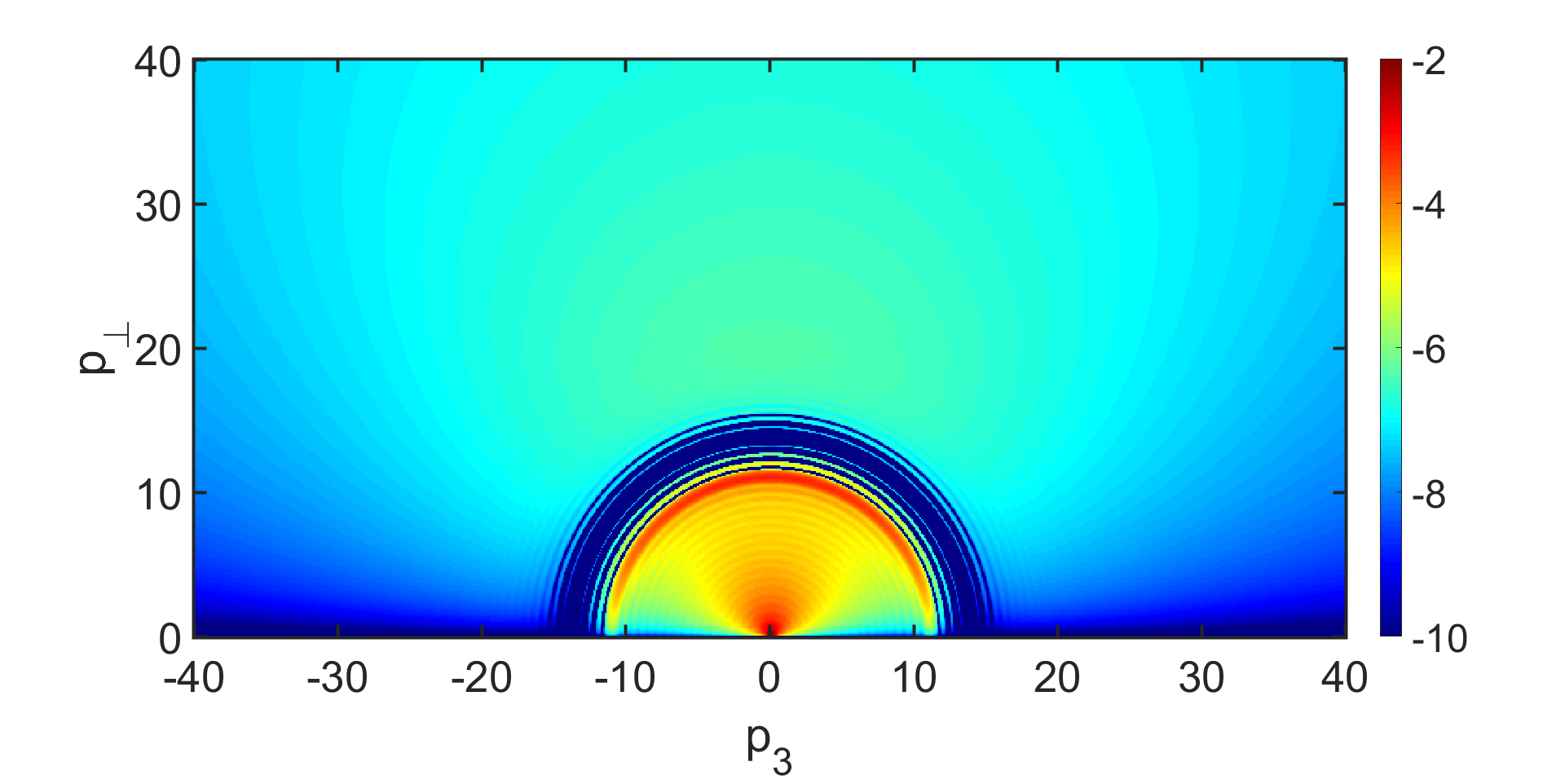}}\
  \subfloat[t=30.0]{
  \includegraphics[width=0.45\linewidth]{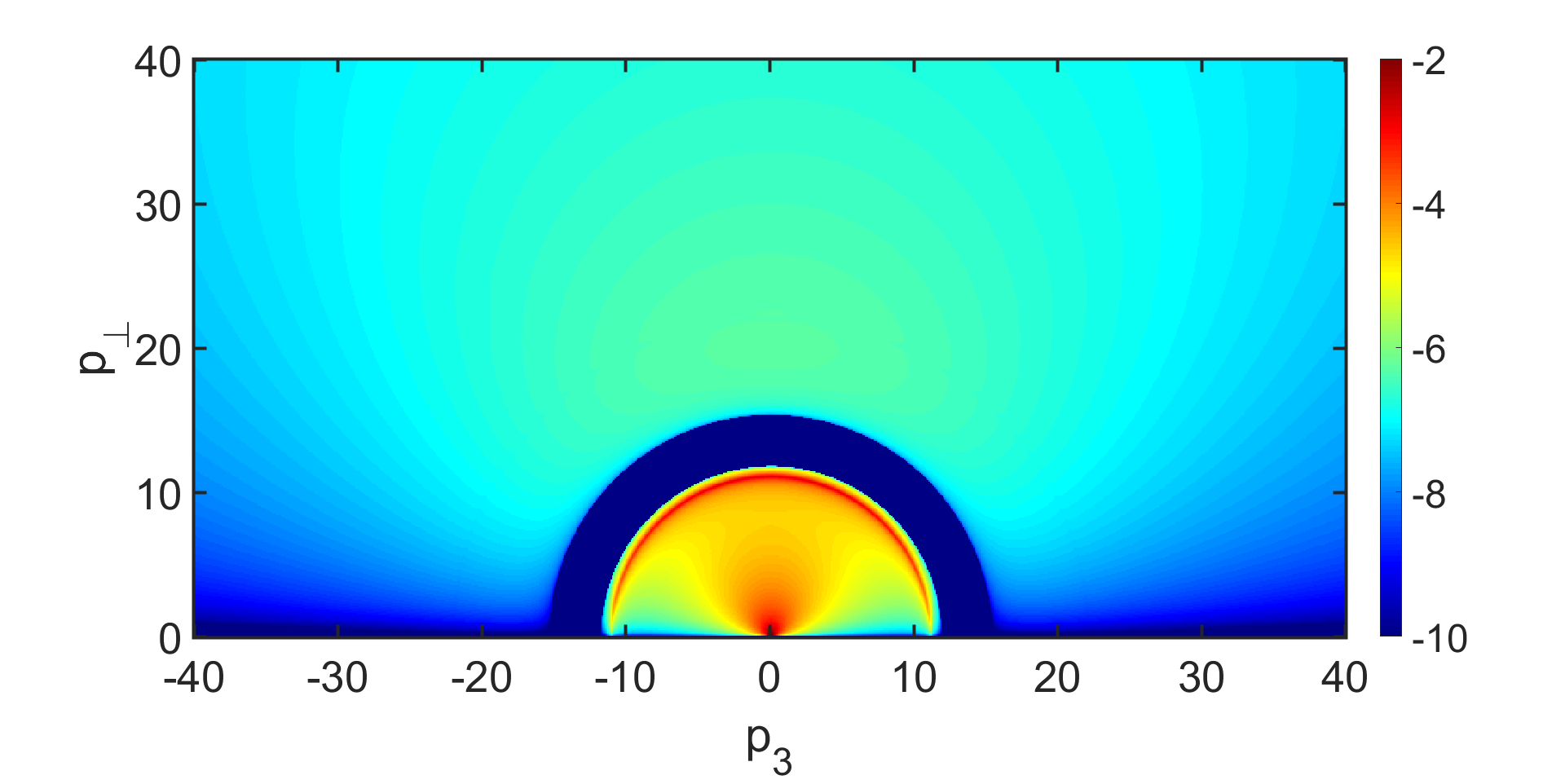}}\\
  \caption{(color on line)The phenomenological distribution function  $Log(f_{br}(t))$ in back reaction field at different time points, where both $p_{\perp}$ and $p_{3}$ are in unit of $\frac{|E_{0}|}{\sqrt{E_{0}}}$}
  \label{fig4}
\end{figure}

In Fig.\ref{fig3} and Fig.\ref{fig4}, we plot the momentum dependence of the distribution function in the Bjorken expanding field and the back reaction field. In Fig.\ref{fig3}, in the Bjorken expanding field, the distribution function concentrates near the region of $p=0$, and symmetrically distribute along $p_{3}=0$. And, distribution function decreases as the momentum increases. At different time points, the momentum dependence of the distribution function keeps the same form. In Fig.\ref{fig4}, the distribution function in back reaction field has the same symmetrical properties. But, as the time goes on, a momentum gap appears, and produced quarks distribute both the near $p=0$ region and the near inside gap region. Which forms a confinement phenomenon.

\begin{figure}[H]
  \centering
  \subfloat[t=0.1]{
  \includegraphics[width=0.45\linewidth]{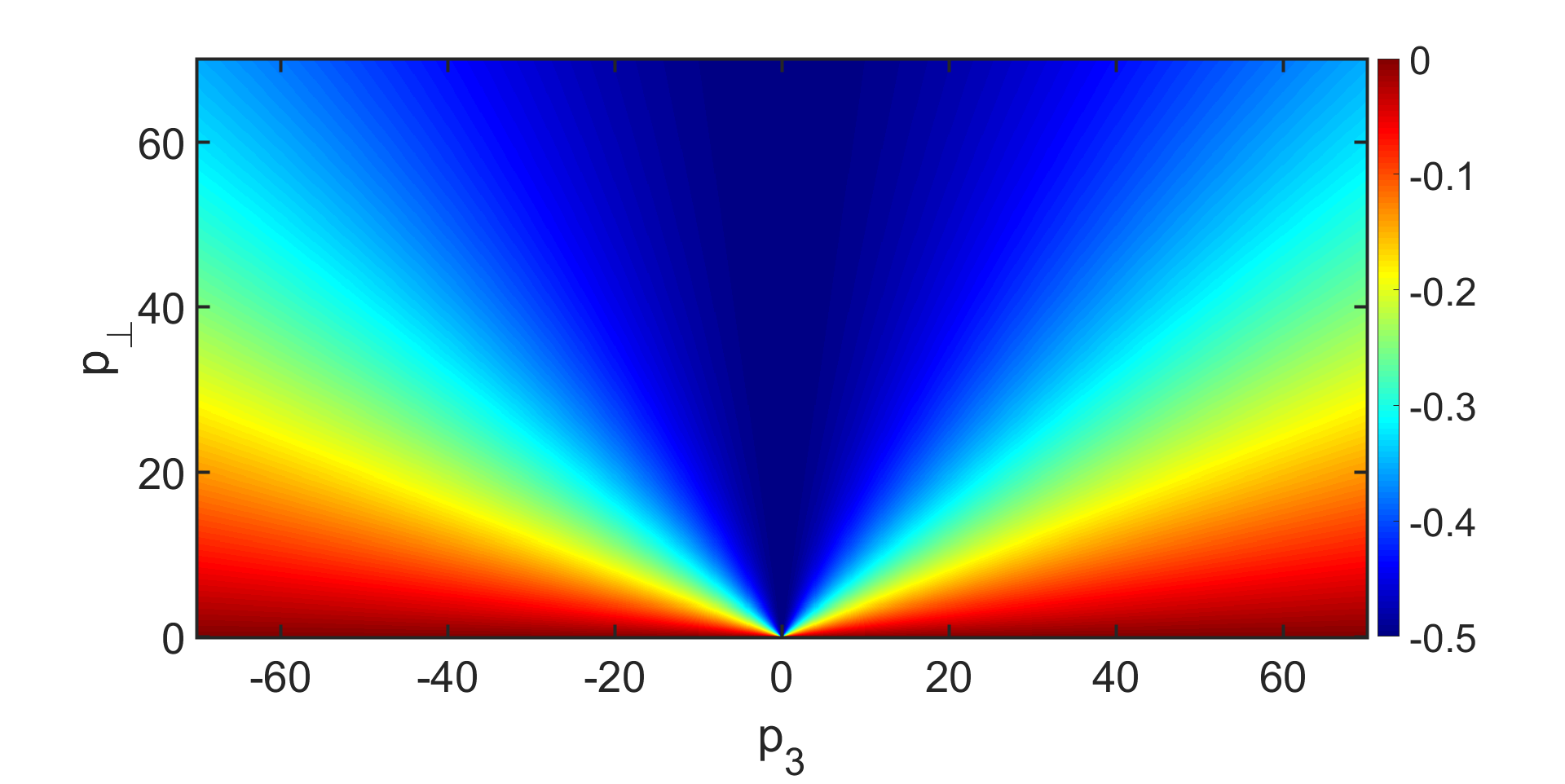}}\
  \subfloat[t=0.3]{
  \includegraphics[width=0.45\linewidth]{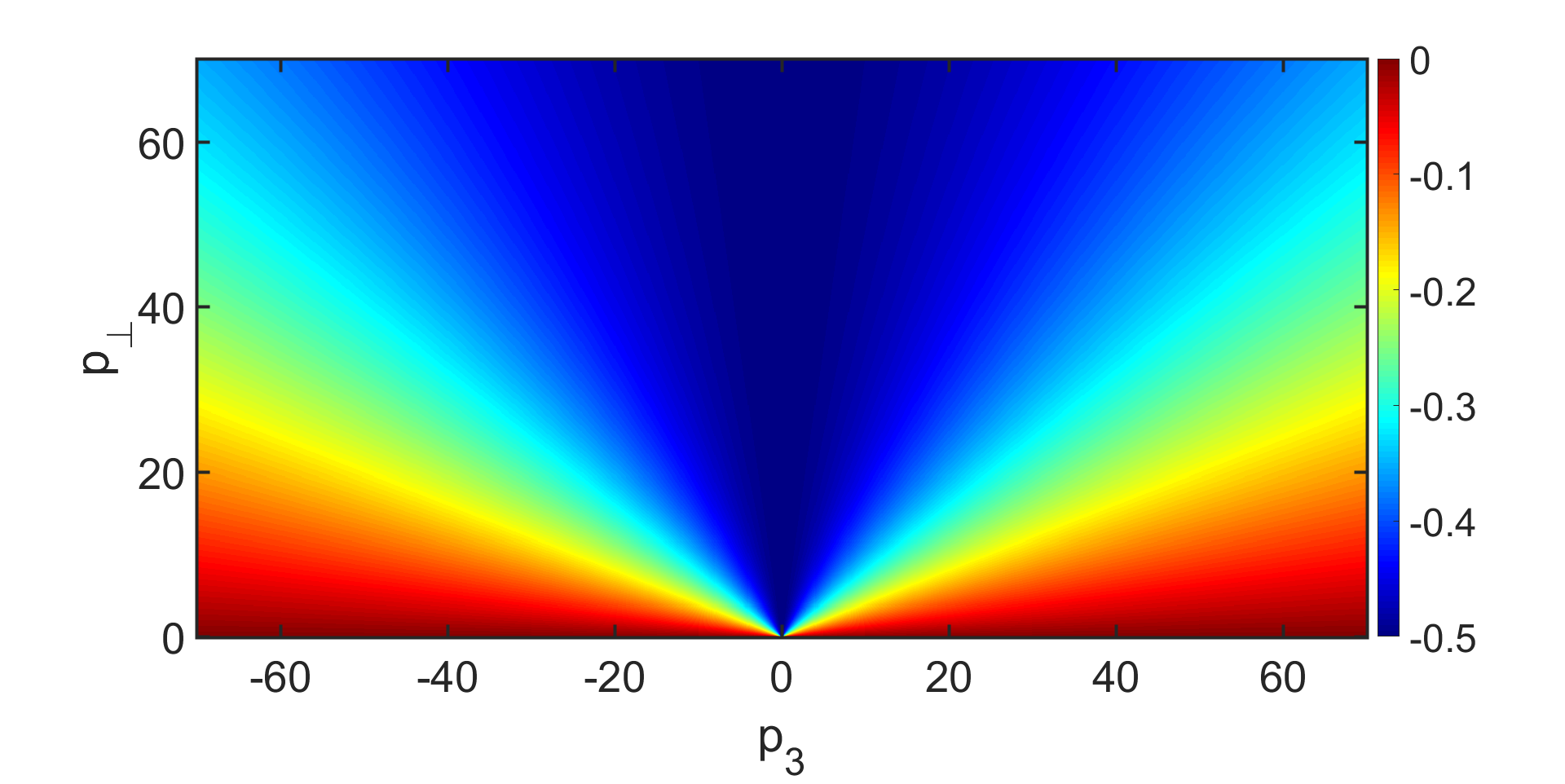}}\\
  \subfloat[t=0.5]{
  \includegraphics[width=0.45\linewidth]{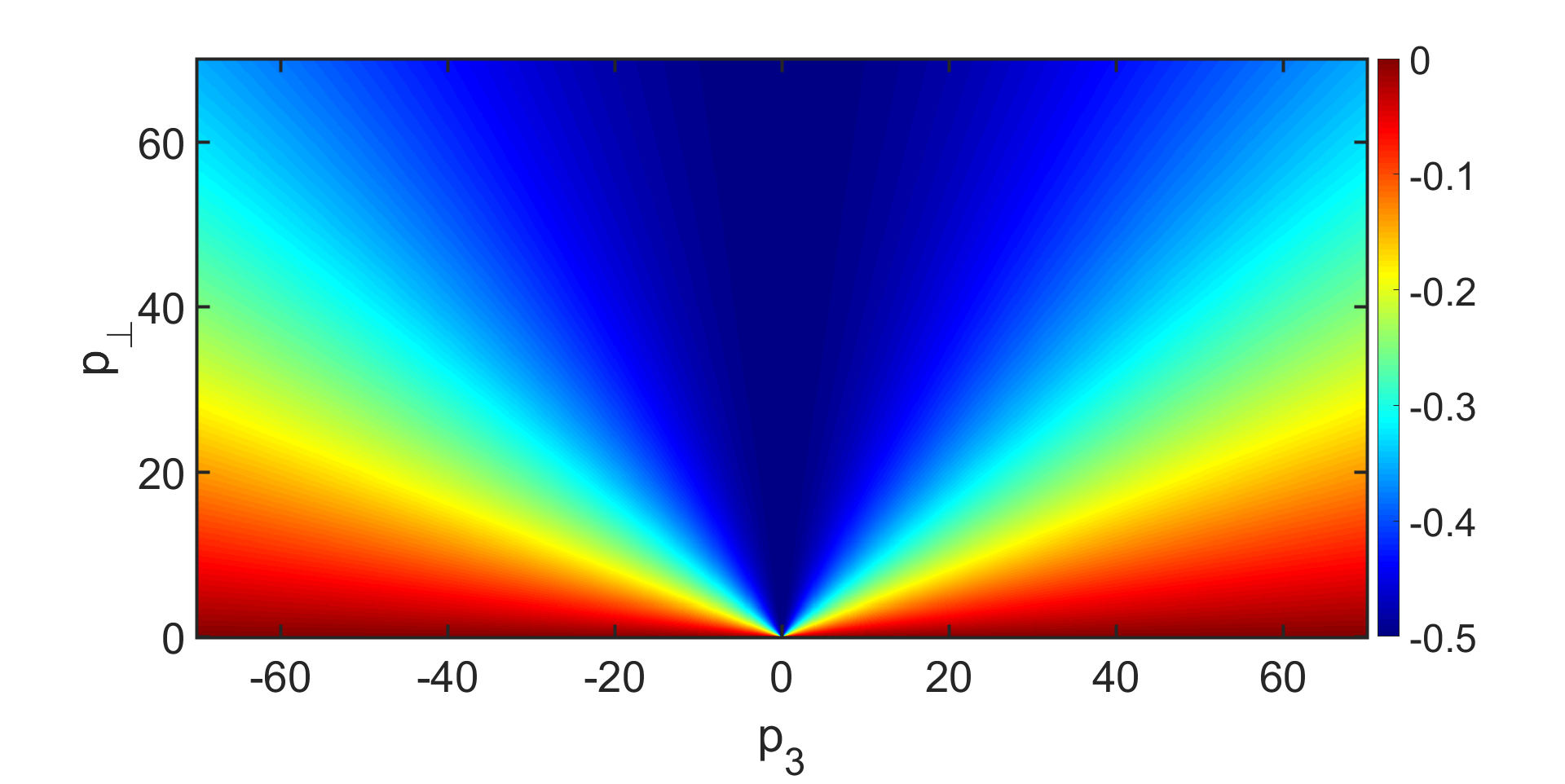}}\
  \subfloat[t=0.7]{
  \includegraphics[width=0.45\linewidth]{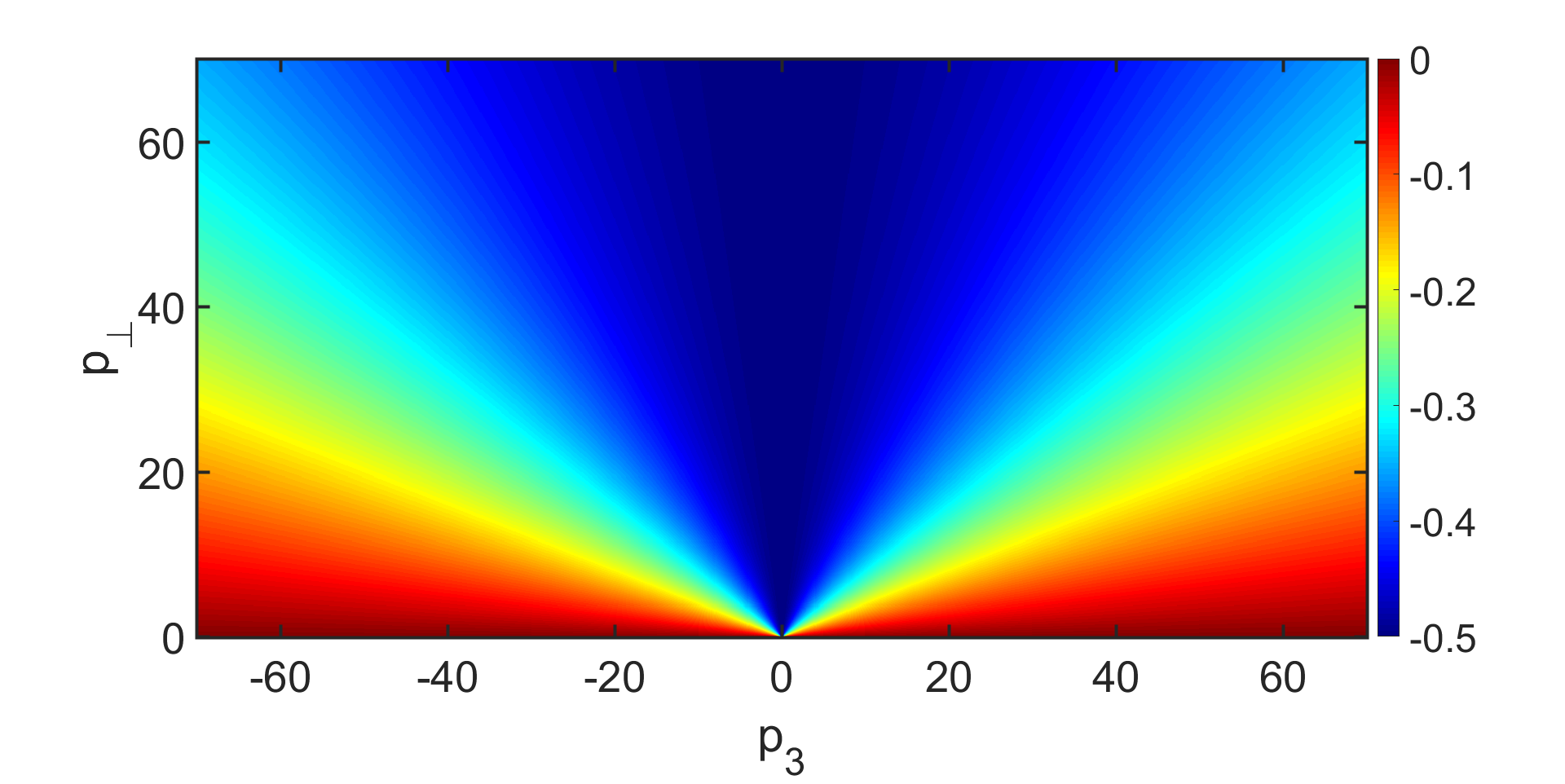}}\\
  \subfloat[t=0.9]{
  \includegraphics[width=0.45\linewidth]{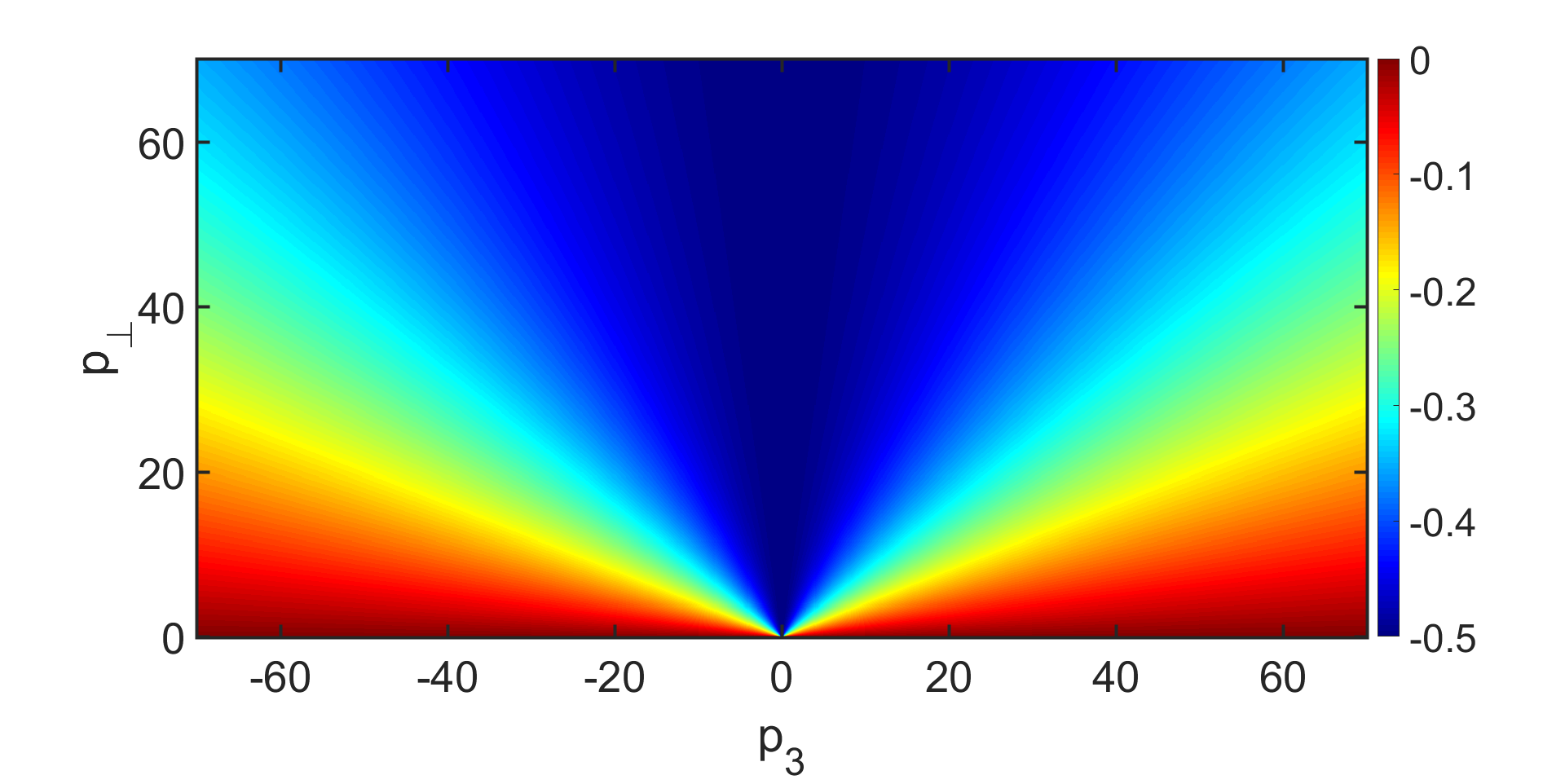}}\
  \subfloat[t=2.0]{
  \includegraphics[width=0.45\linewidth]{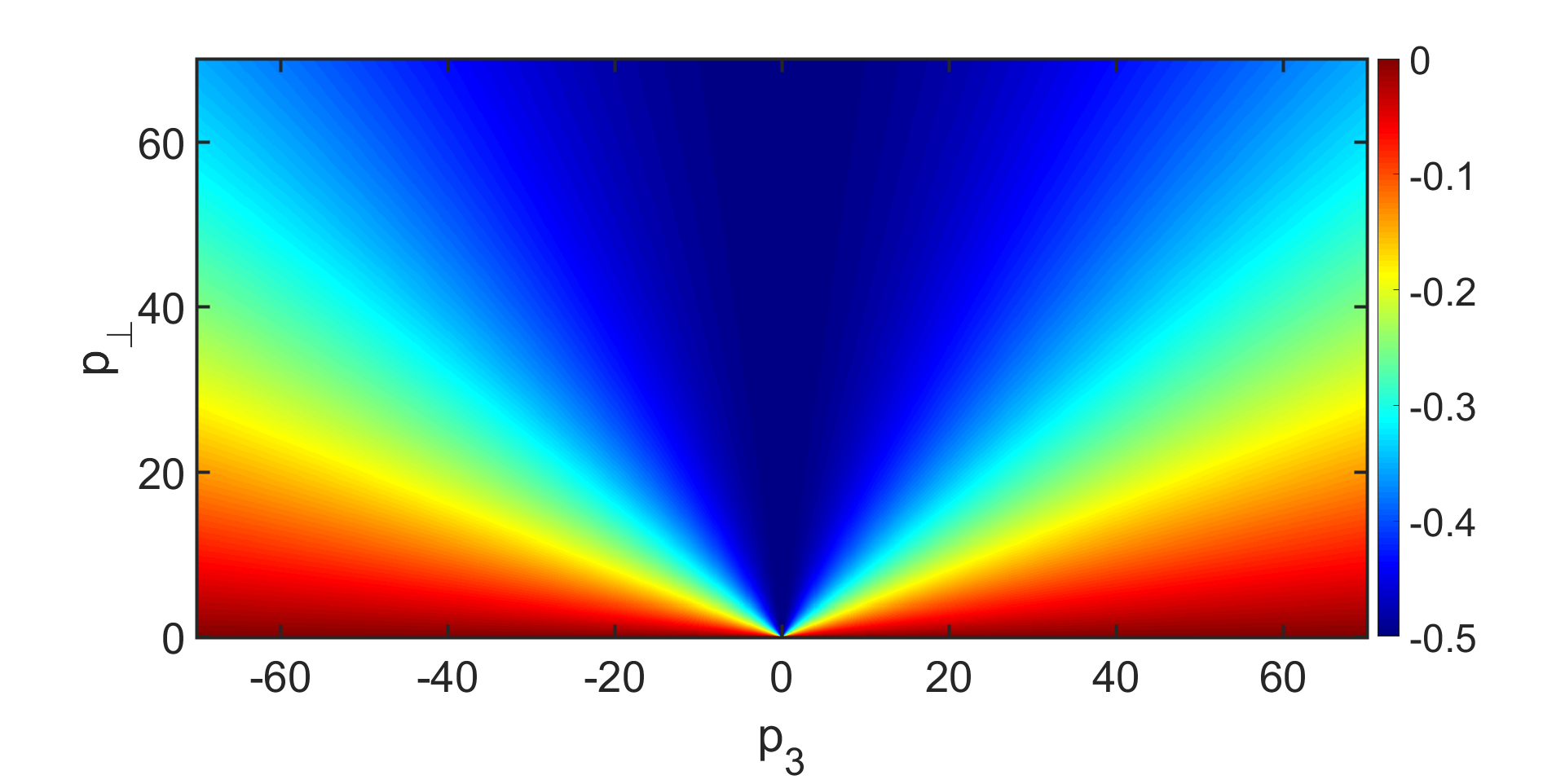}}\\
  \subfloat[t=5.0]{
  \includegraphics[width=0.45\linewidth]{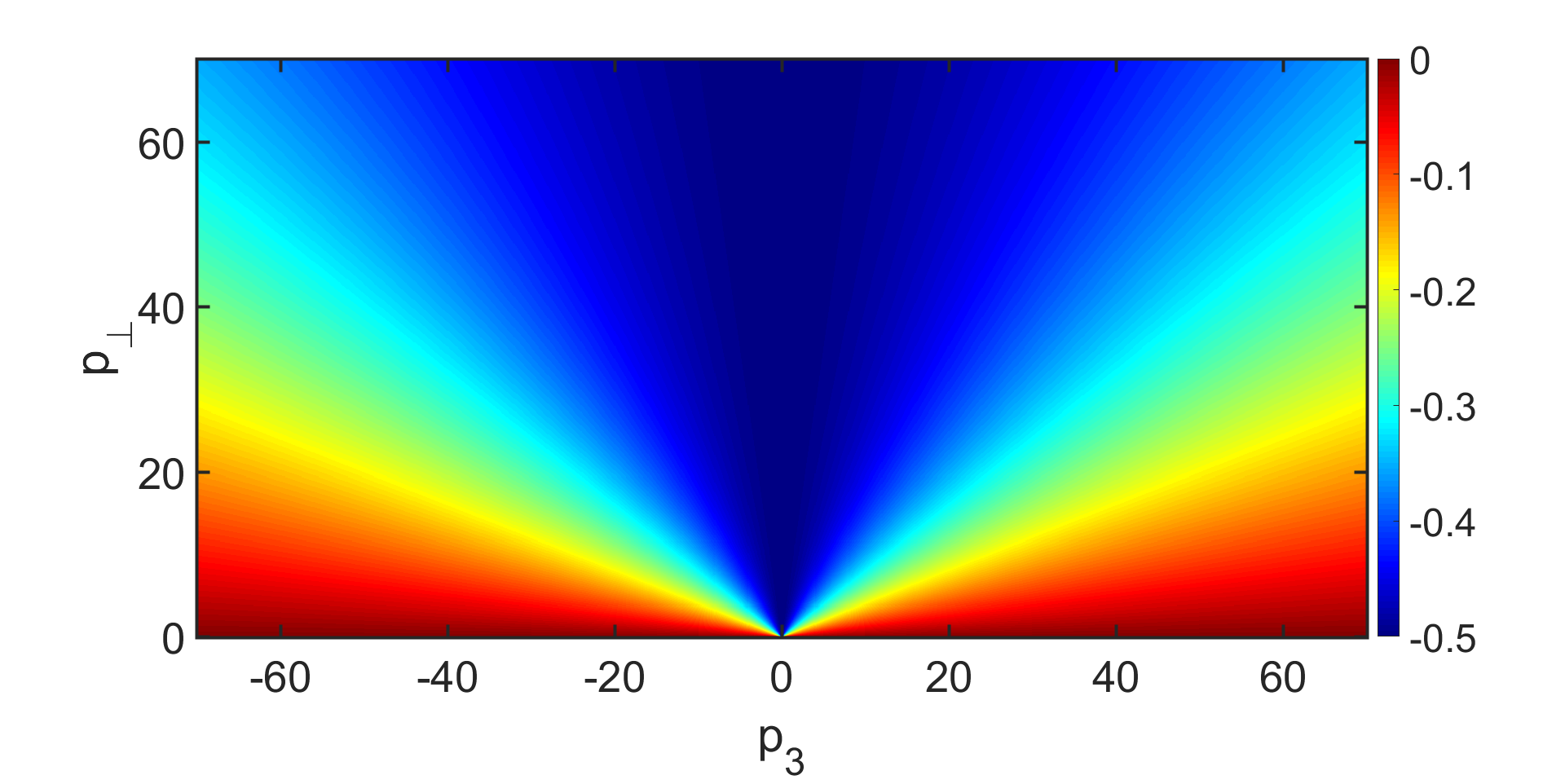}}\
  \subfloat[t=30.0]{
  \includegraphics[width=0.45\linewidth]{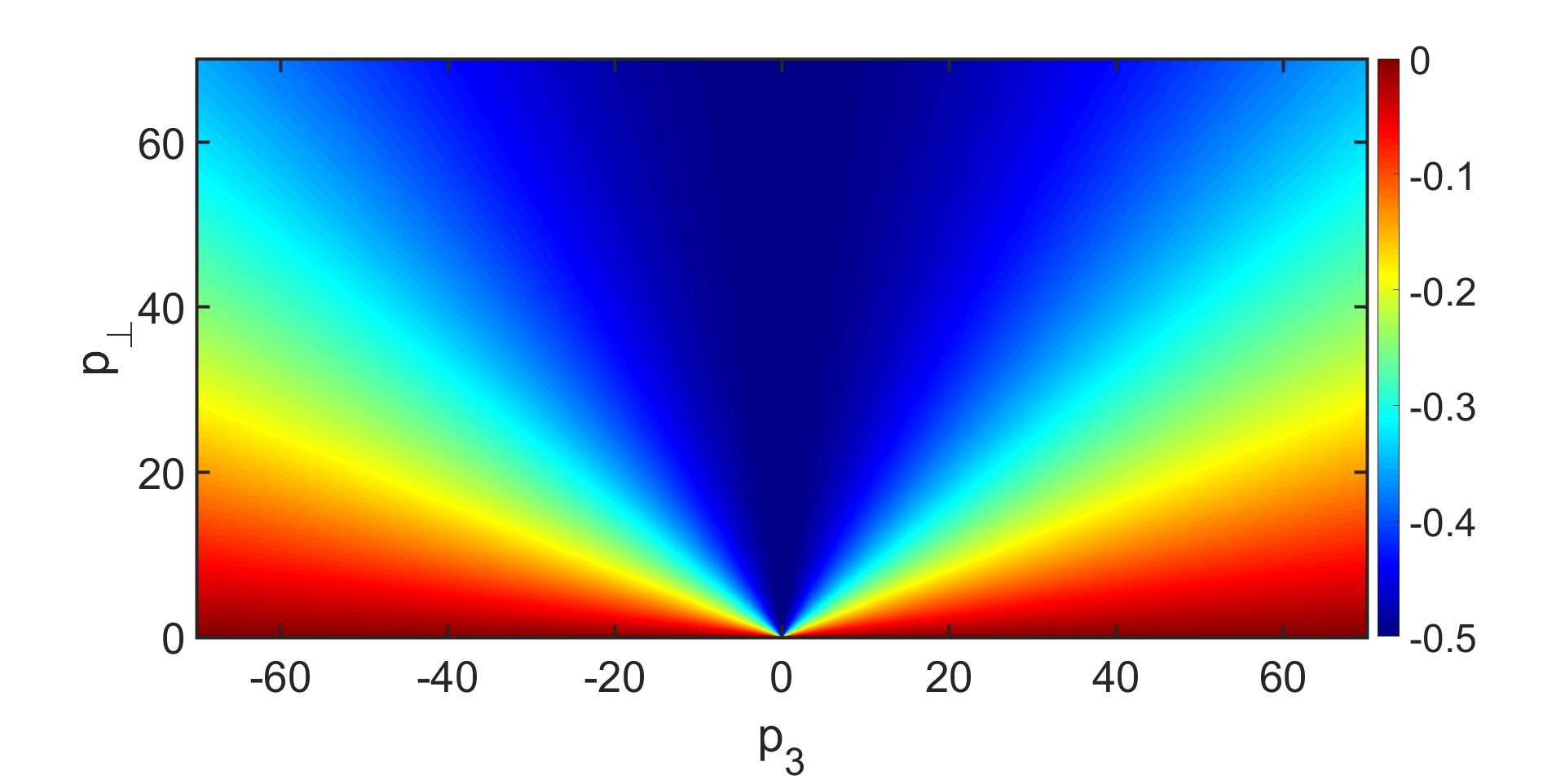}}\\
  \caption{(color on line) $b_{\perp}^{s}(p_{\perp},p_{3})$ at different times, where both $p_{\perp}$ and $p_{3}$ are in unit of $\frac{|E_{0}|}{\sqrt{E_{0}}}$}
  \label{fig5}
\end{figure}

\begin{figure}[H]
  \centering
  \subfloat[t=0.1]{
  \includegraphics[width=0.45\linewidth]{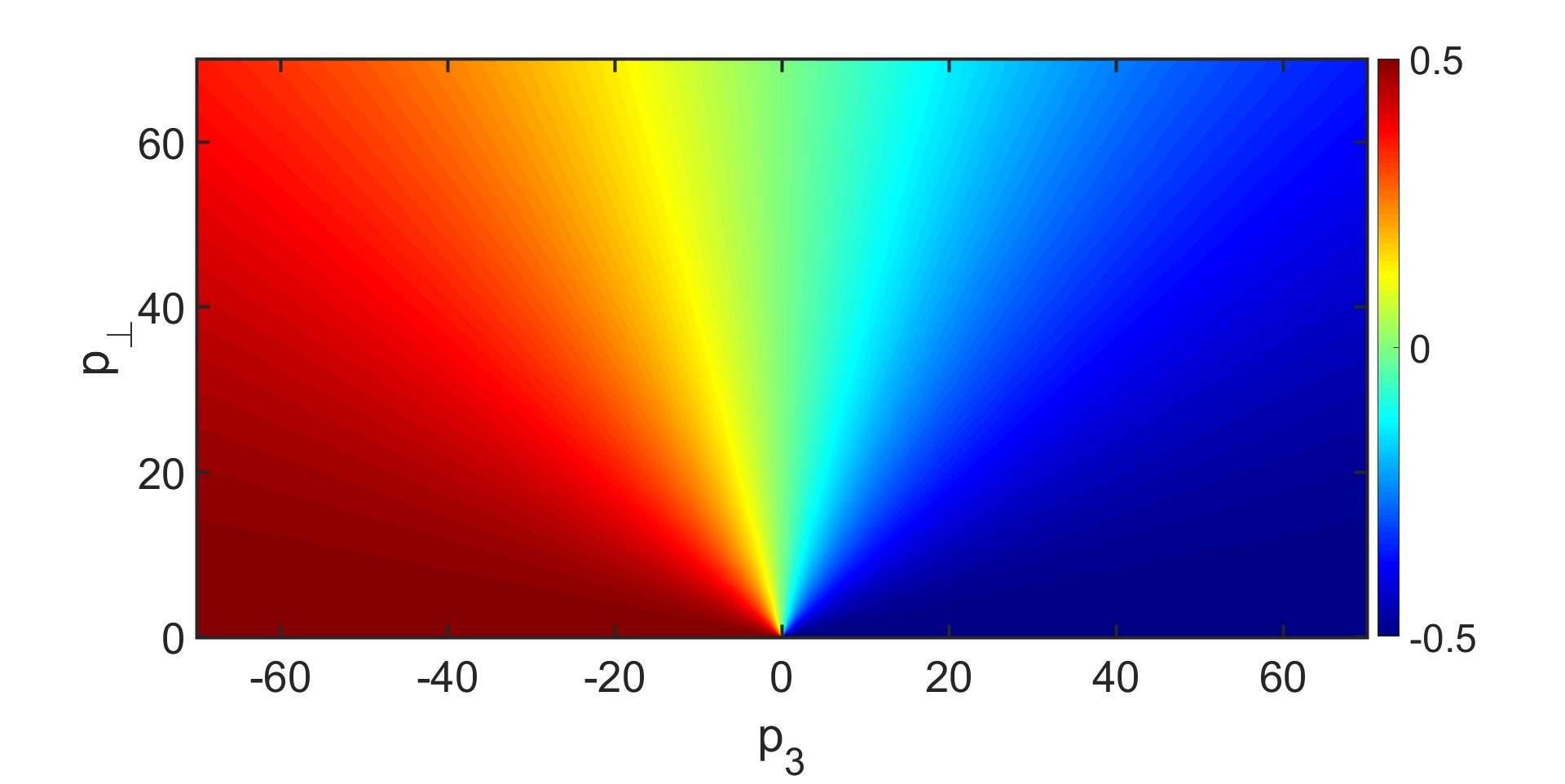}}\
  \subfloat[t=0.3]{
  \includegraphics[width=0.45\linewidth]{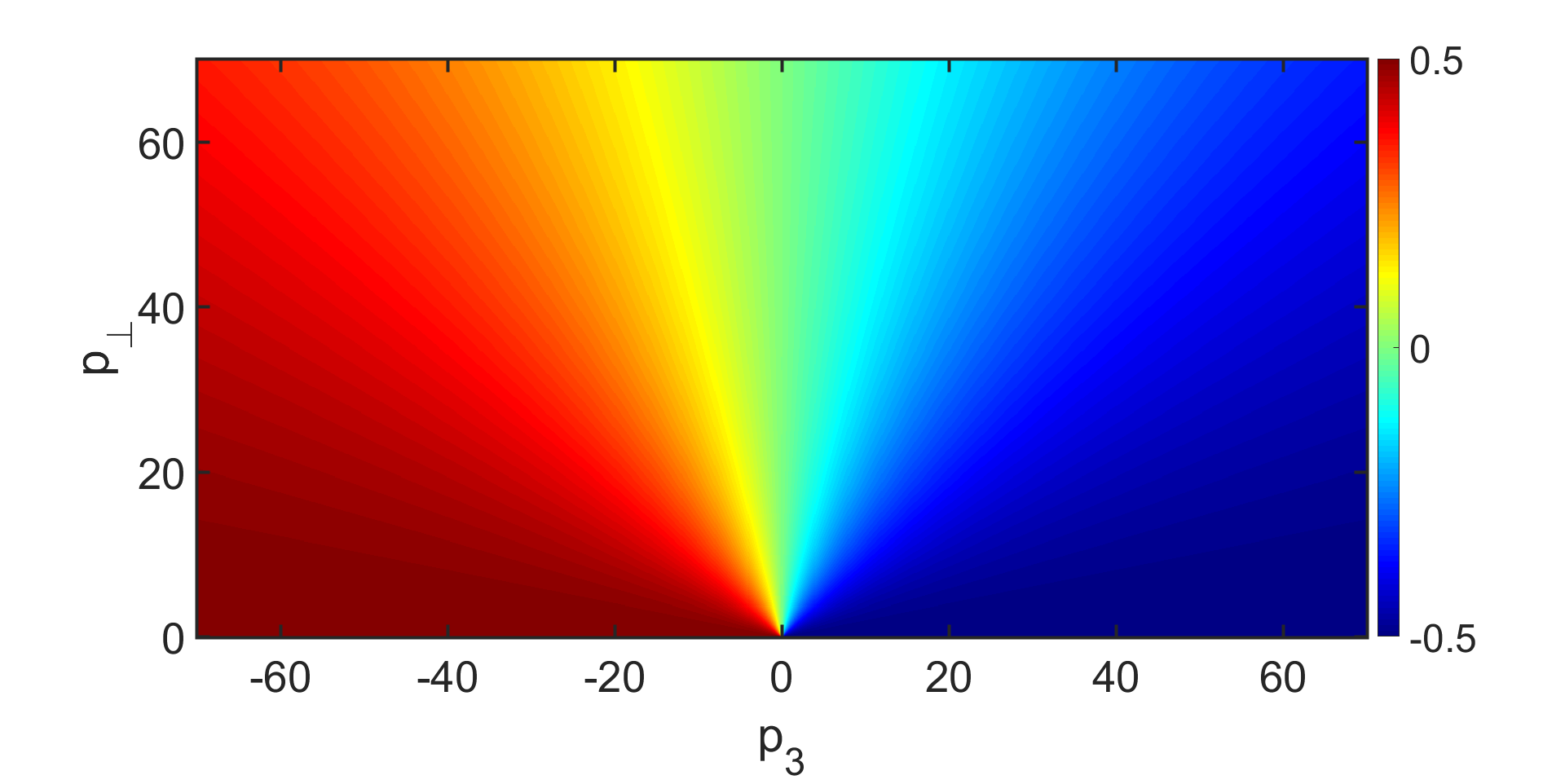}}\\
  \subfloat[t=0.5]{
  \includegraphics[width=0.45\linewidth]{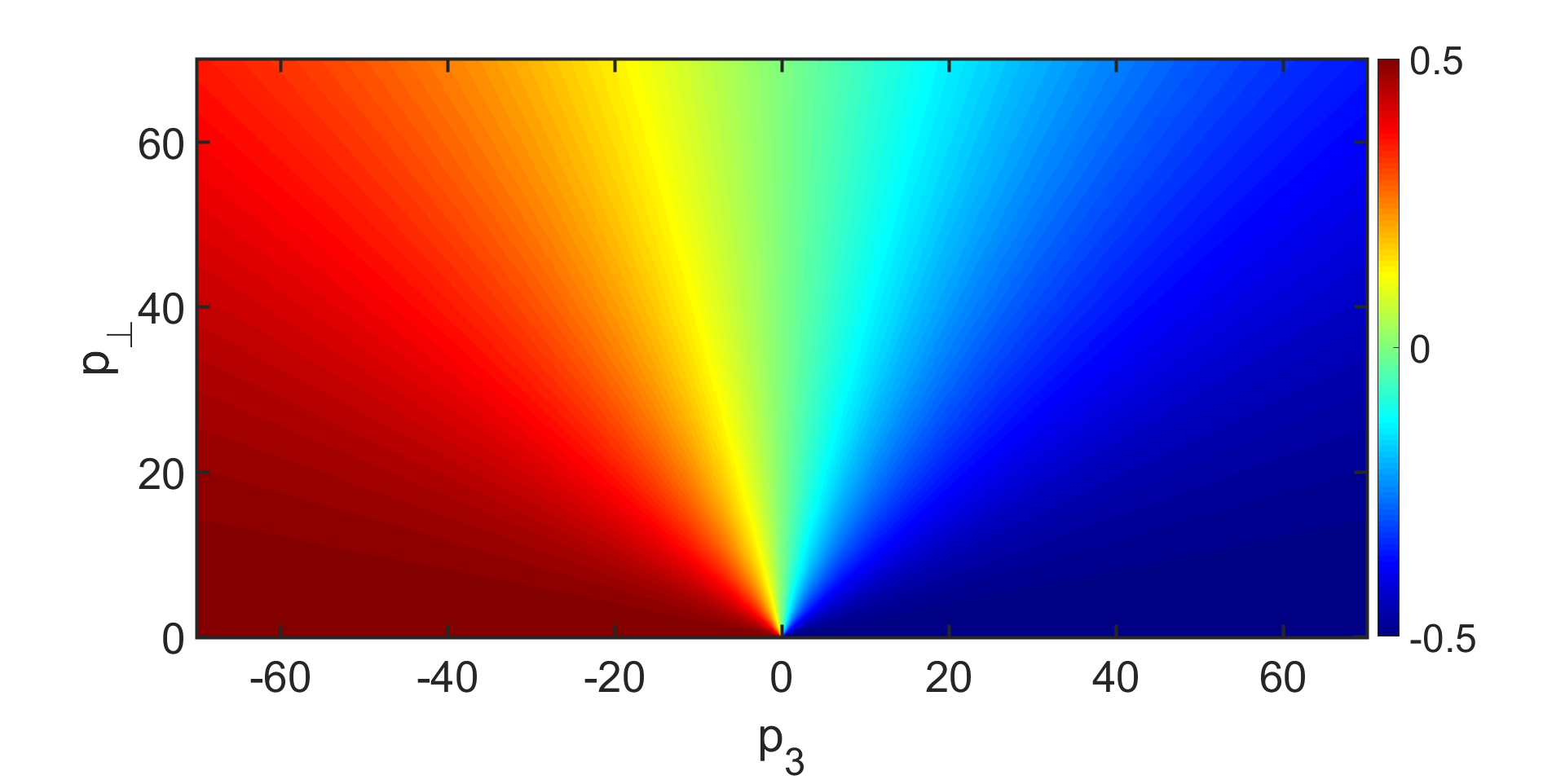}}\
  \subfloat[t=0.7]{
  \includegraphics[width=0.45\linewidth]{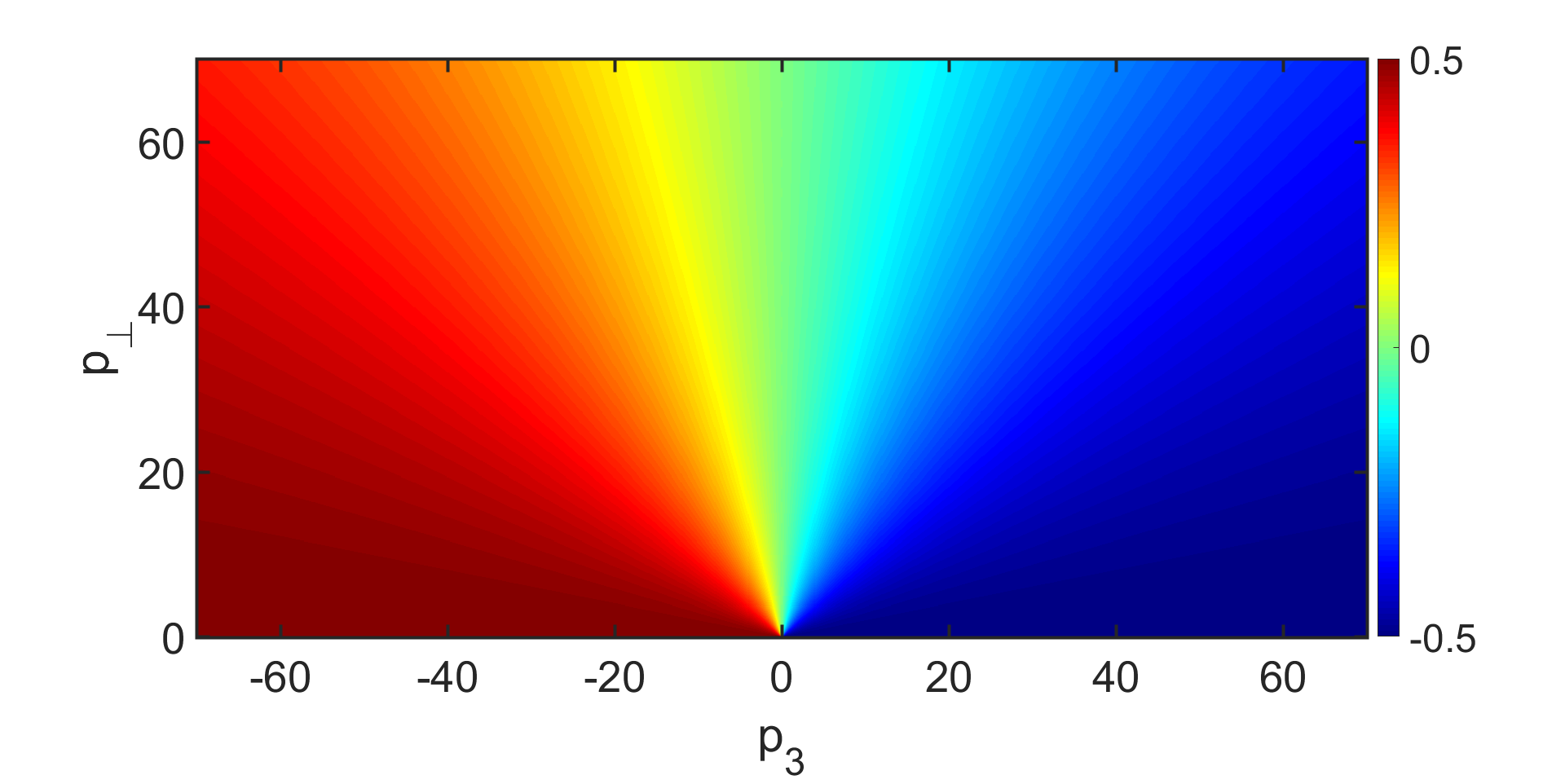}}\\
  \subfloat[t=0.9]{
  \includegraphics[width=0.45\linewidth]{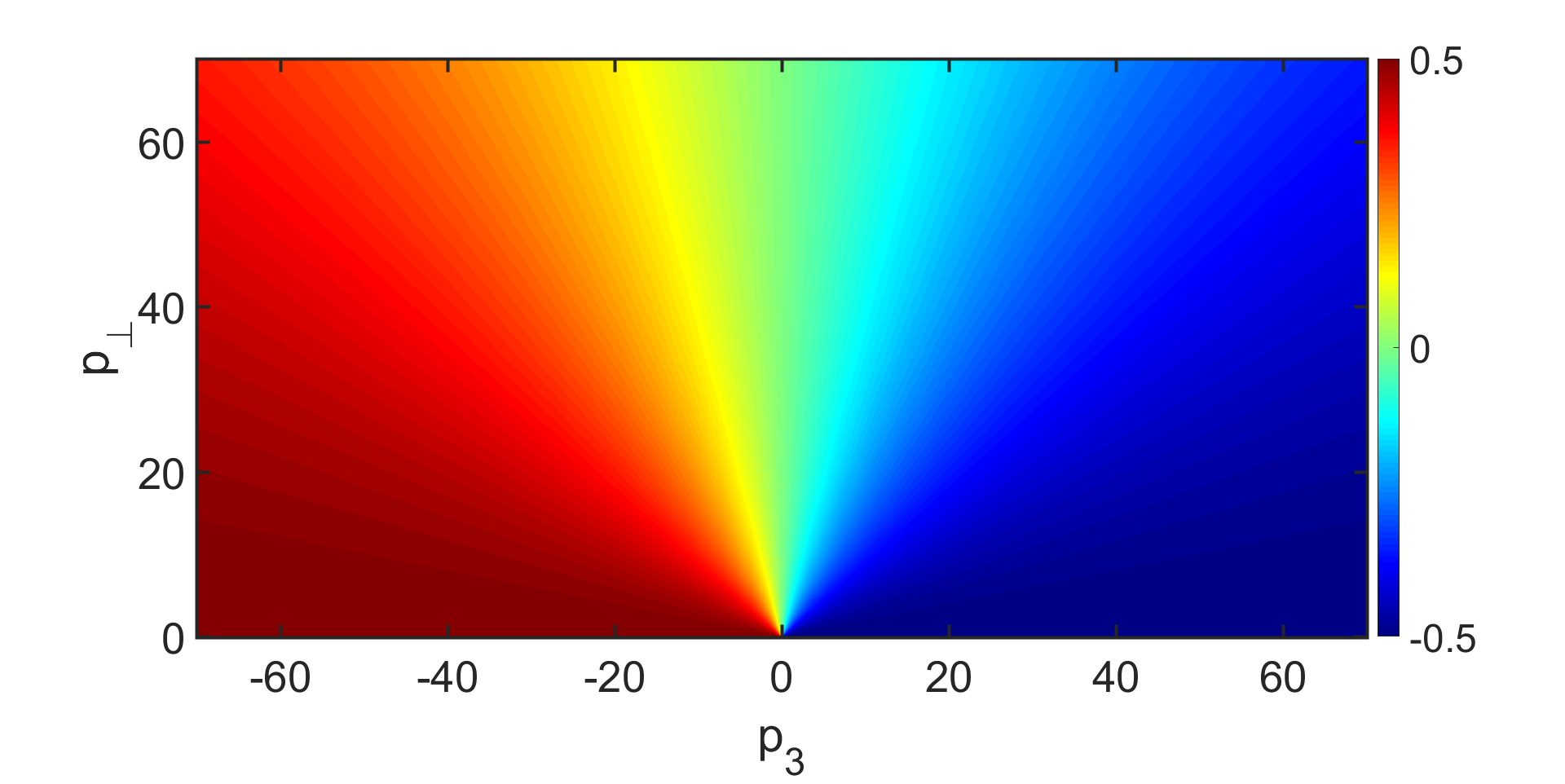}}\
  \subfloat[t=2.0]{
  \includegraphics[width=0.45\linewidth]{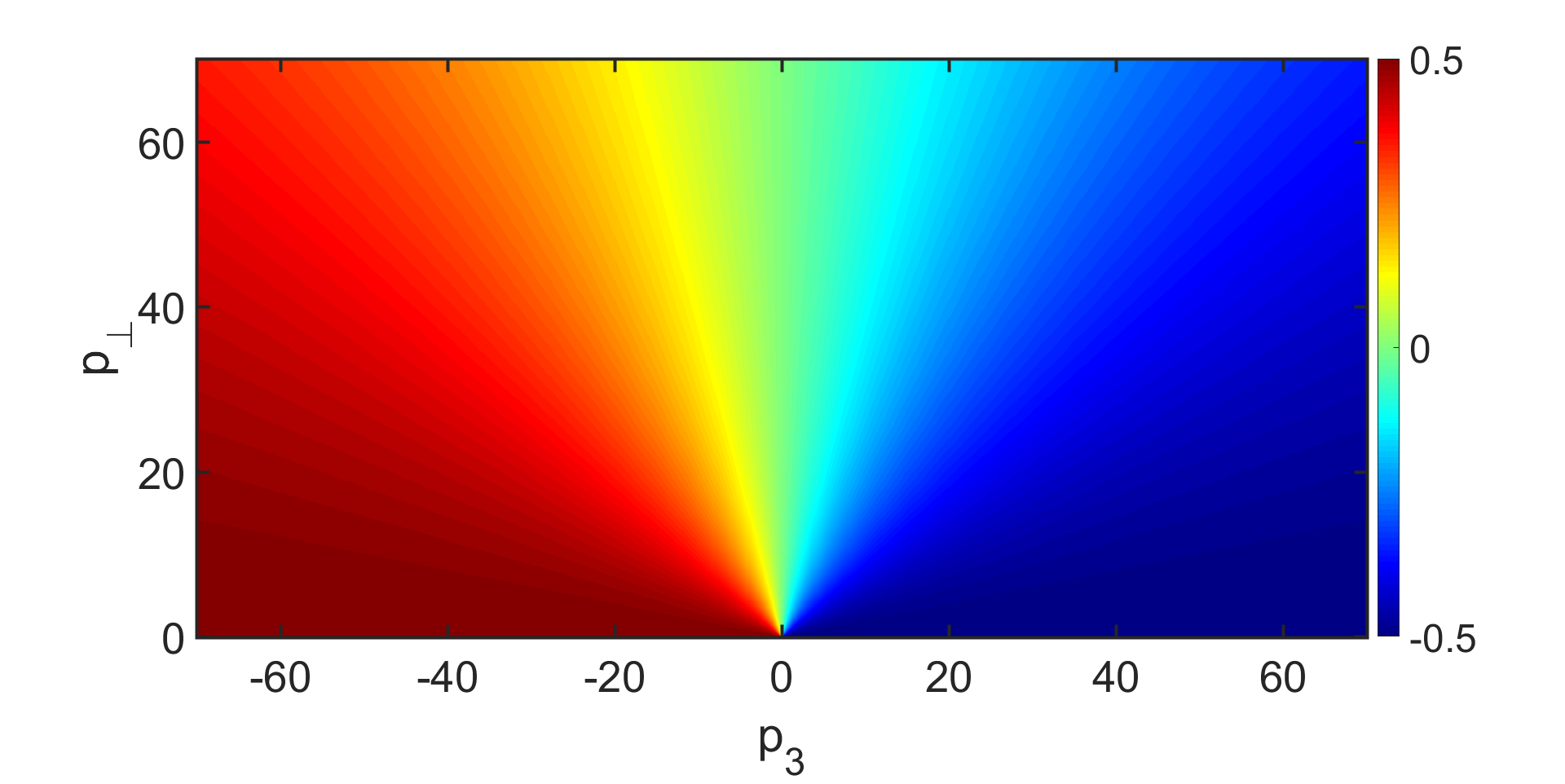}}\\
  \subfloat[t=5.0]{
  \includegraphics[width=0.45\linewidth]{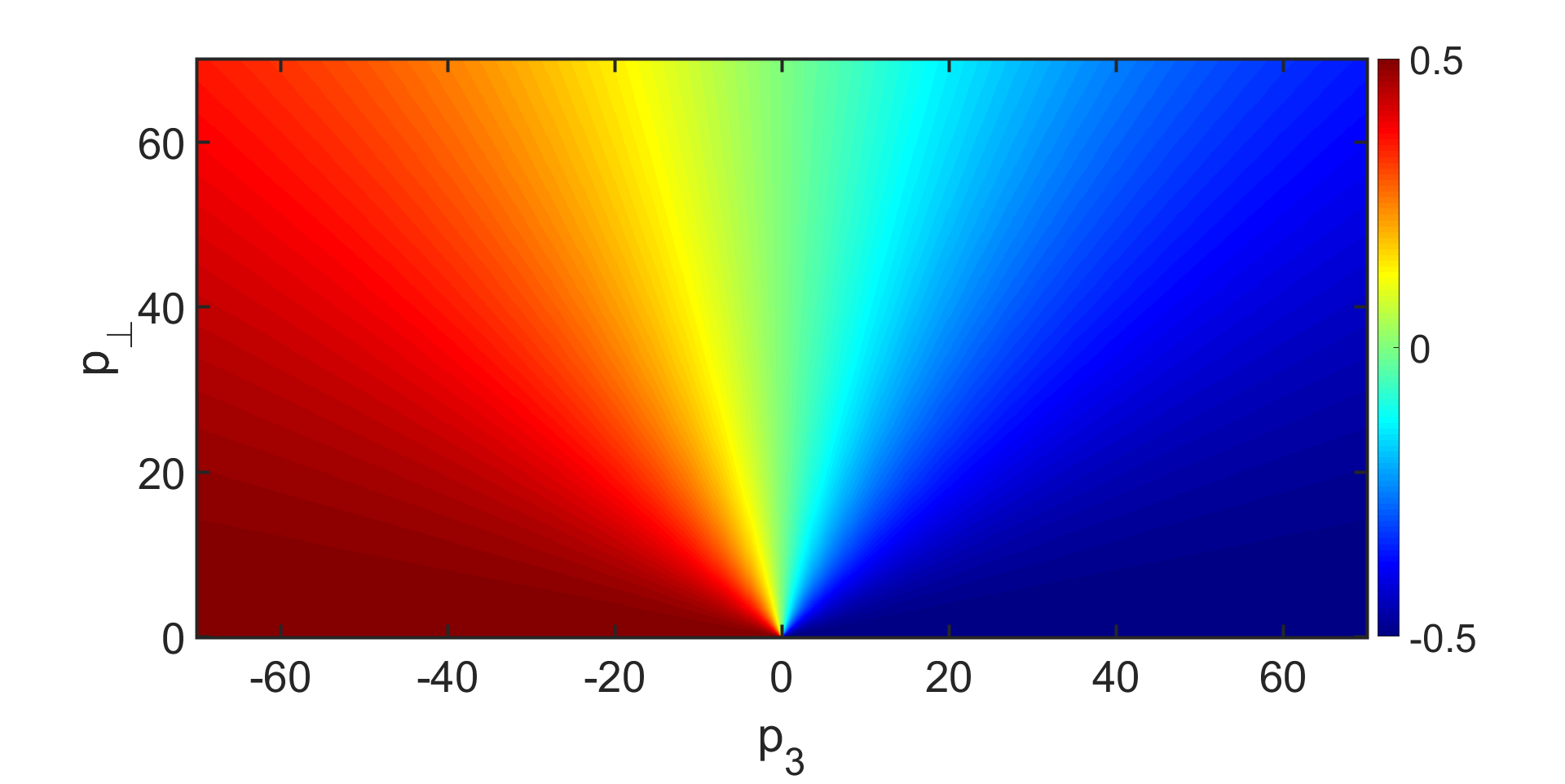}}\
  \subfloat[t=30.0]{
  \includegraphics[width=0.45\linewidth]{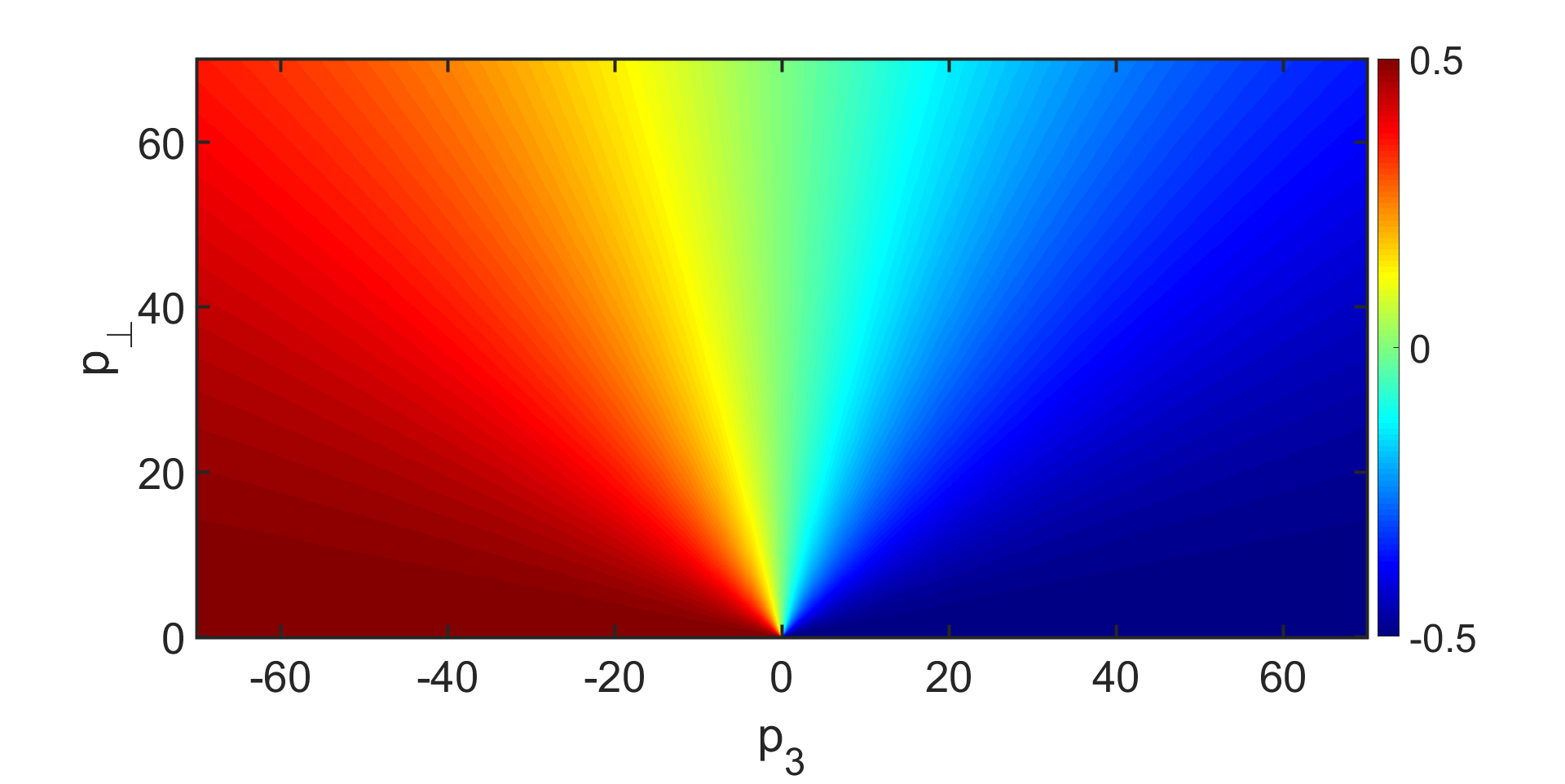}}\\
  \caption{(color on line) $b_{3}^{s}(p_{\perp},p_{3})$ at different times, where both $p_{\perp}$ and $p_{3}$ are in unit of $\frac{|E_{0}|}{\sqrt{E_{0}}}$}
  \label{fig6}
\end{figure}

\begin{figure}[H]
  \centering
  \subfloat[t=0.1]{
  \includegraphics[width=0.45\linewidth]{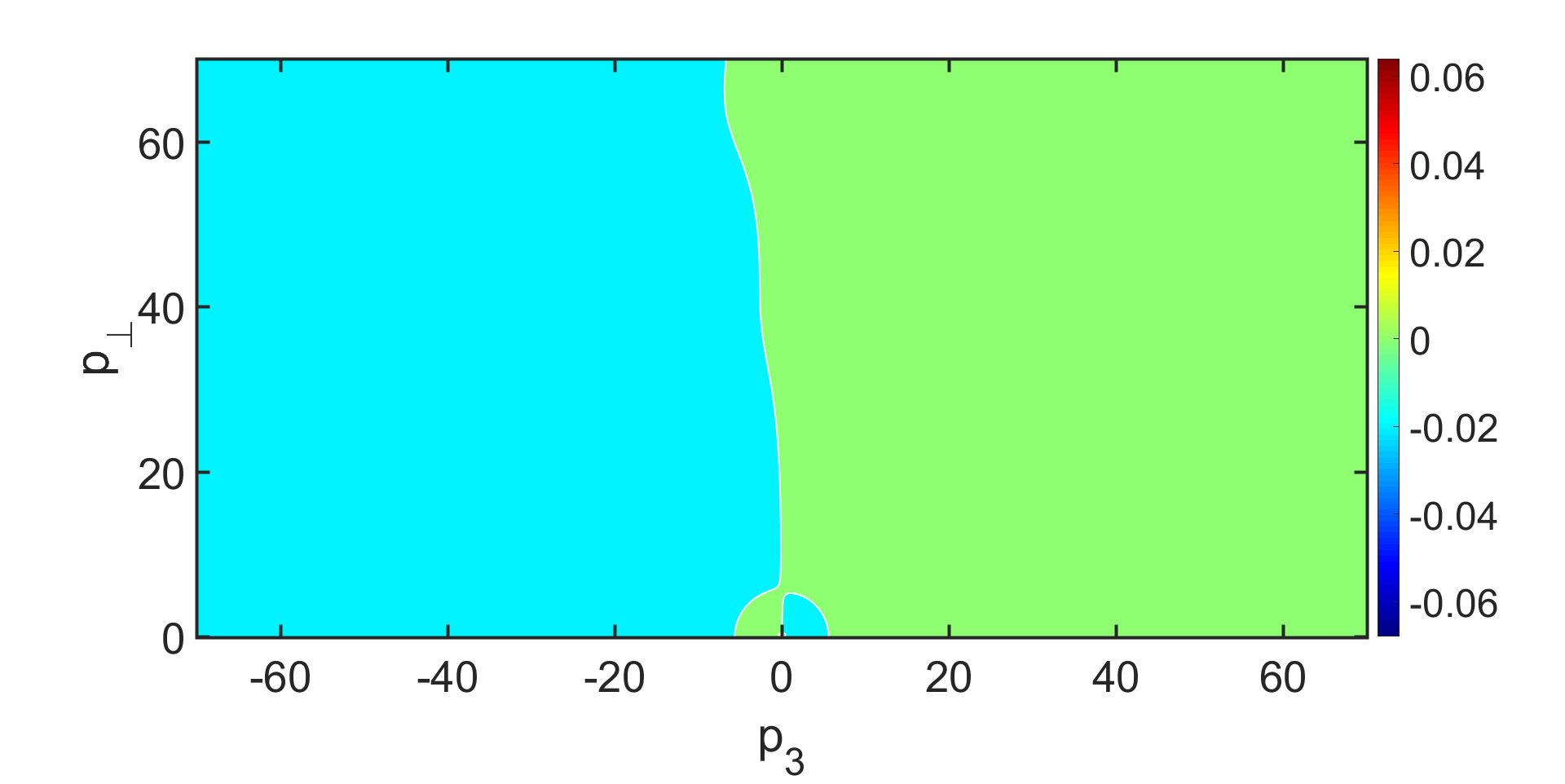}}\
  \subfloat[t=1.0]{
  \includegraphics[width=0.45\linewidth]{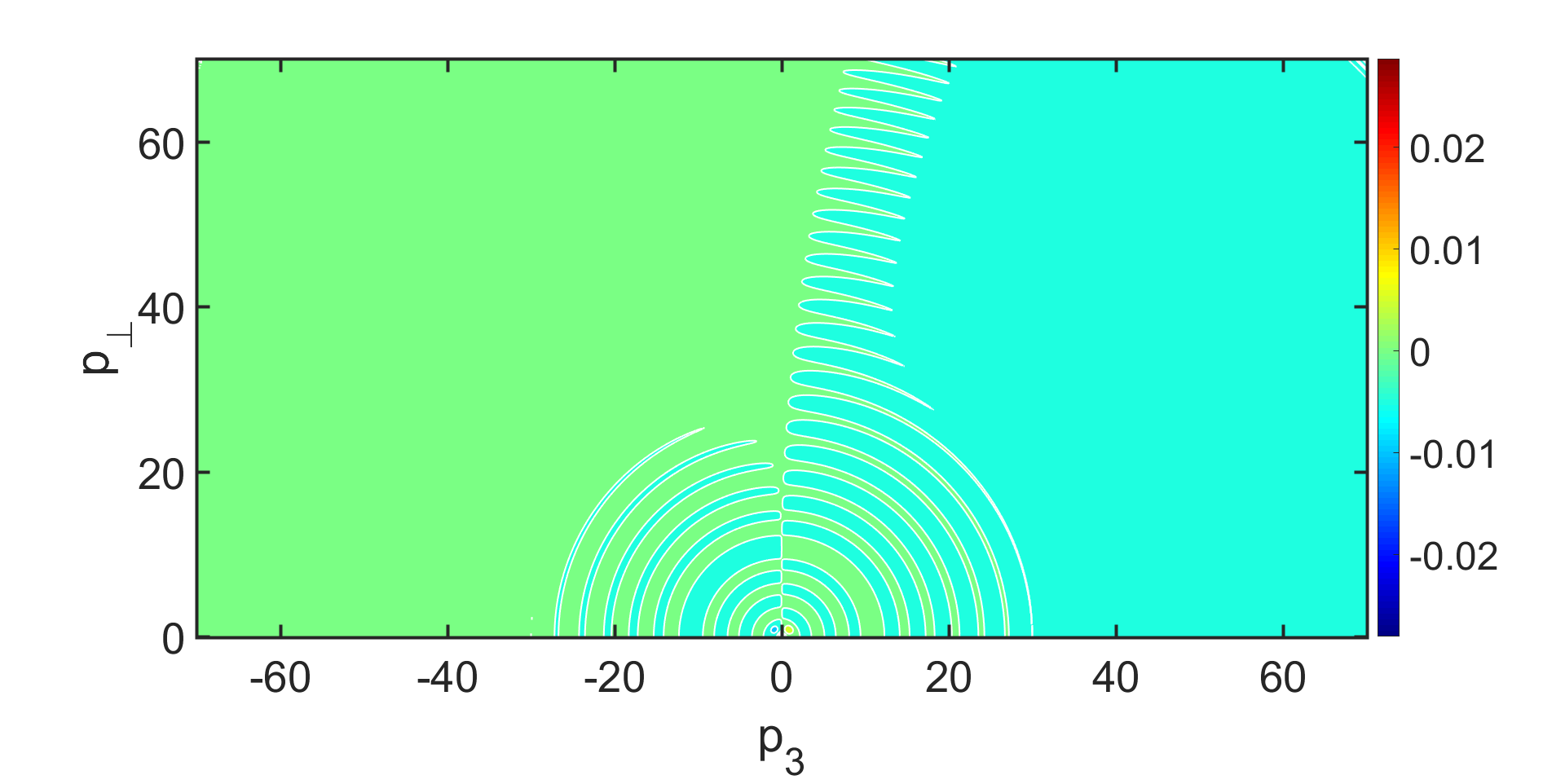}}\\
  \subfloat[t=5.0]{
  \includegraphics[width=0.45\linewidth]{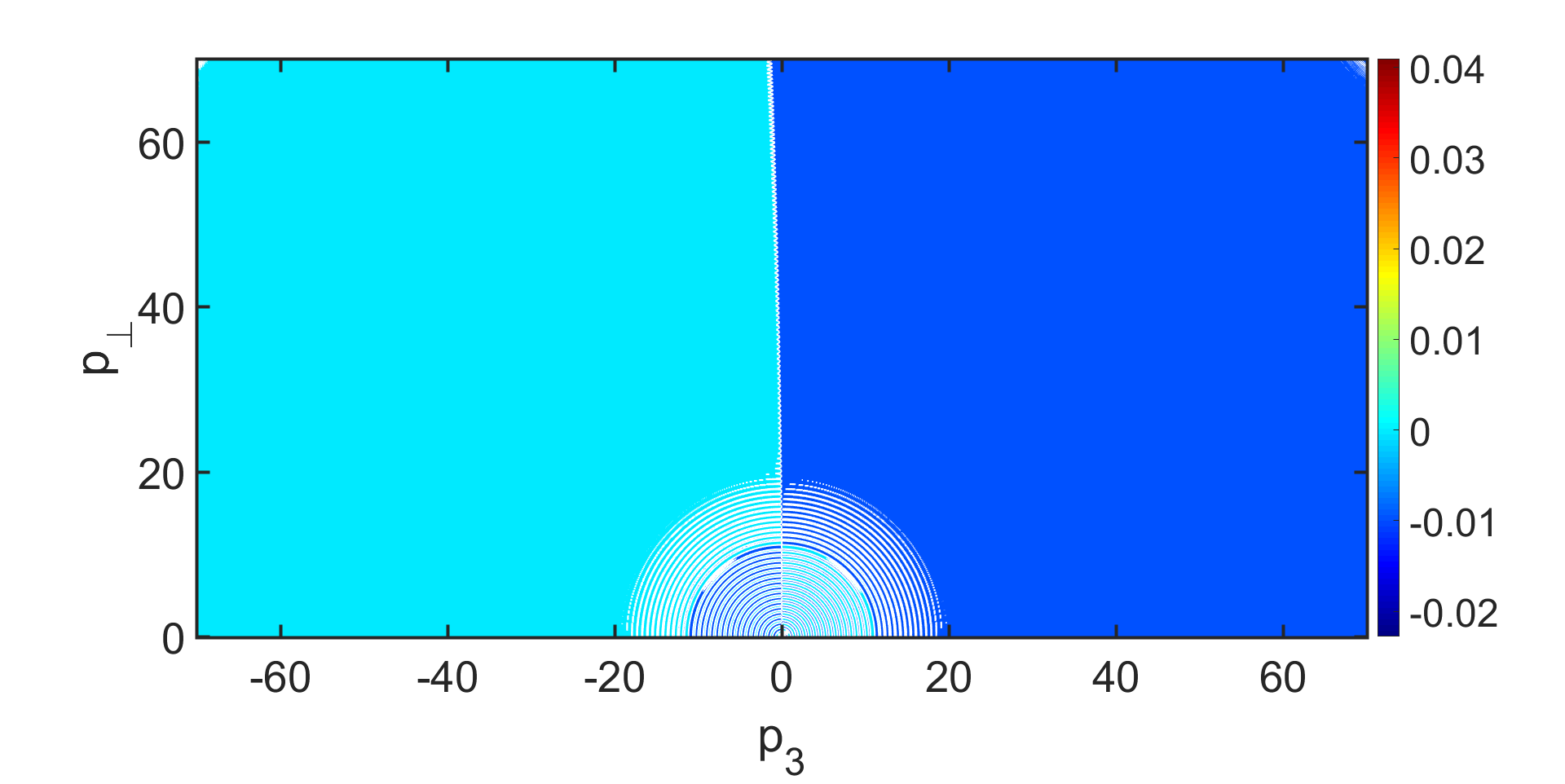}}\
  \subfloat[t=10.0]{
  \includegraphics[width=0.45\linewidth]{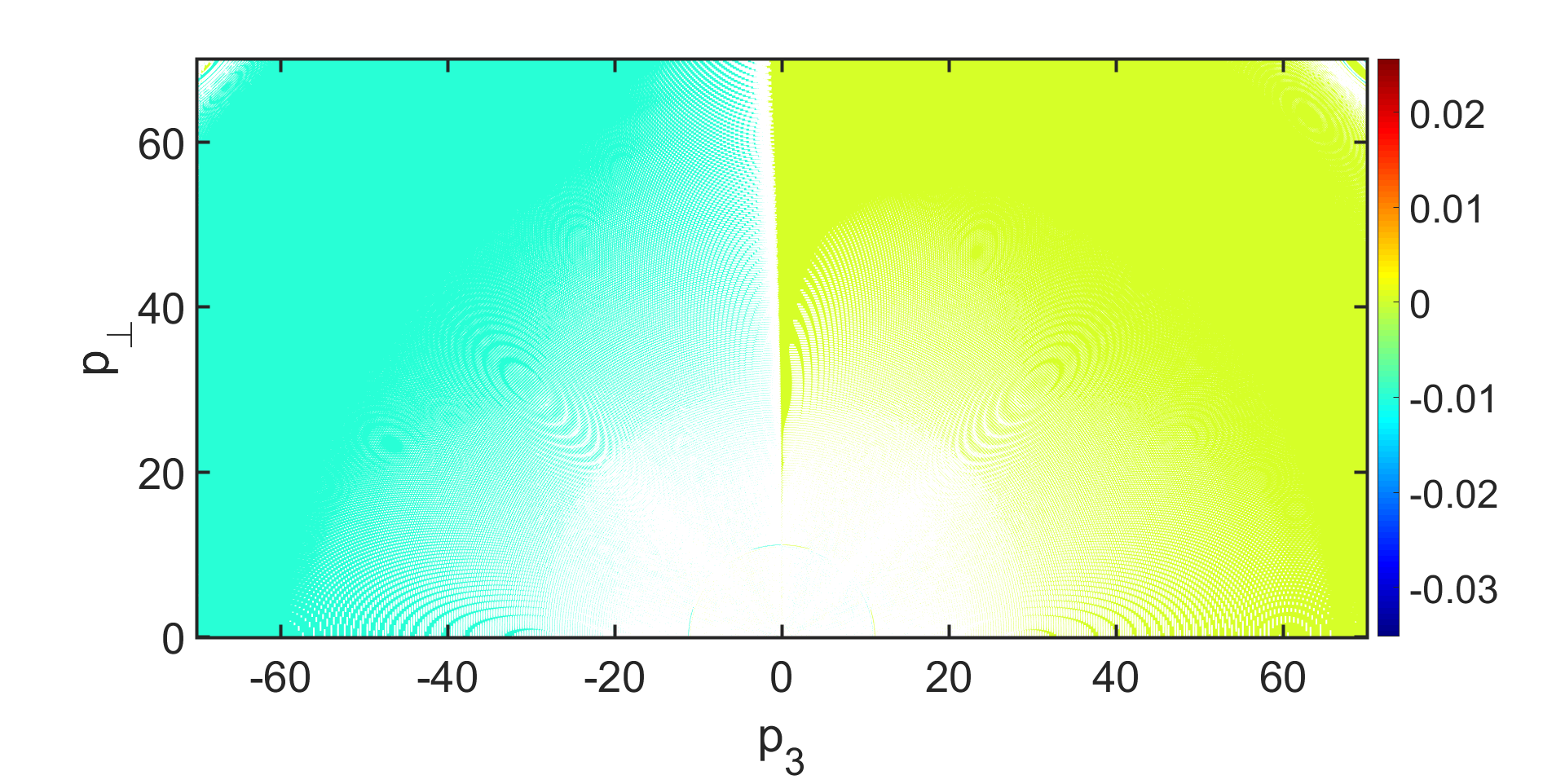}}\\
  \subfloat[t=15.0]{
  \includegraphics[width=0.45\linewidth]{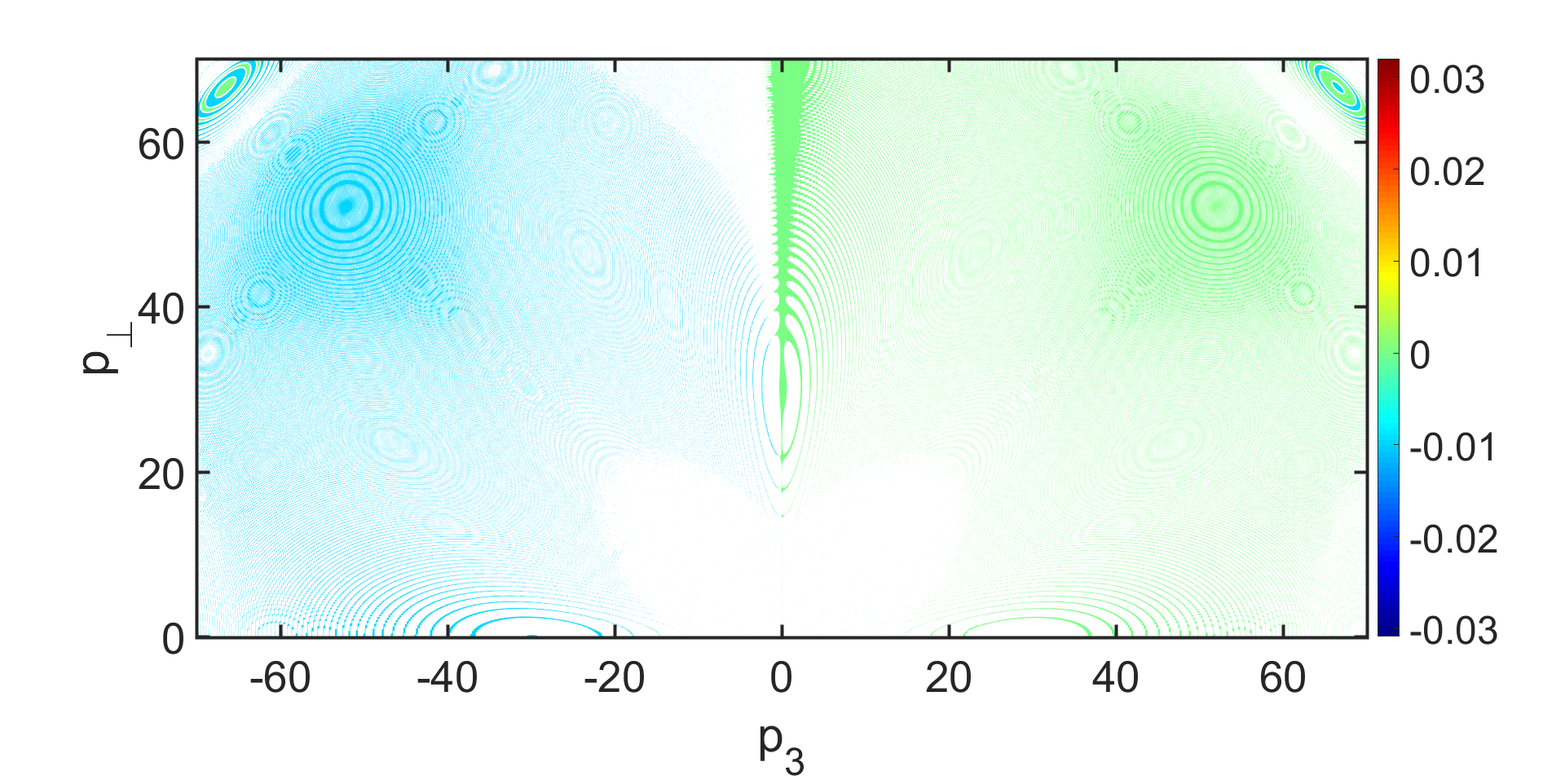}}\
  \subfloat[t=20.0]{
  \includegraphics[width=0.45\linewidth]{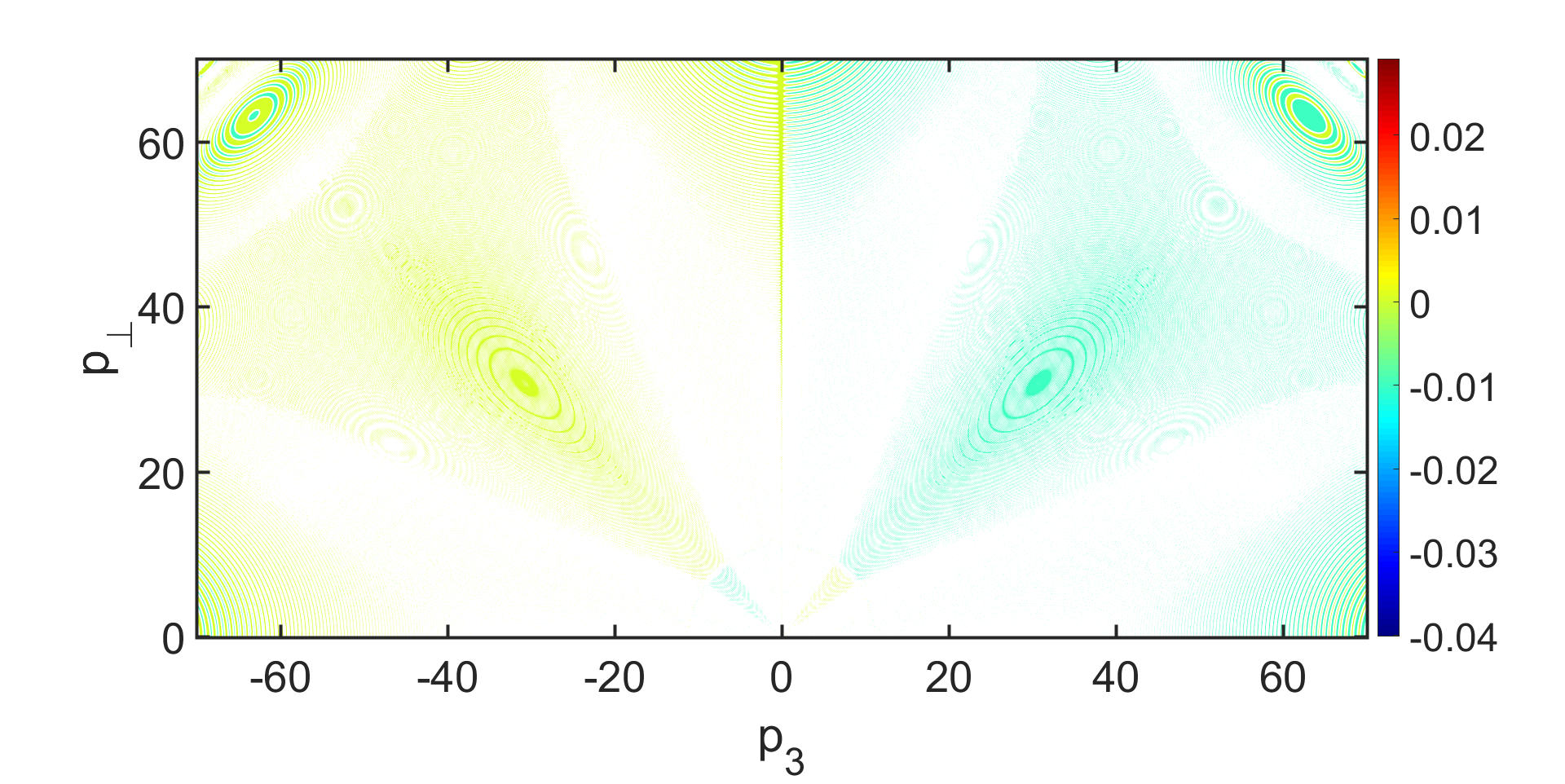}}\\
  \subfloat[t=25.0]{
  \includegraphics[width=0.45\linewidth]{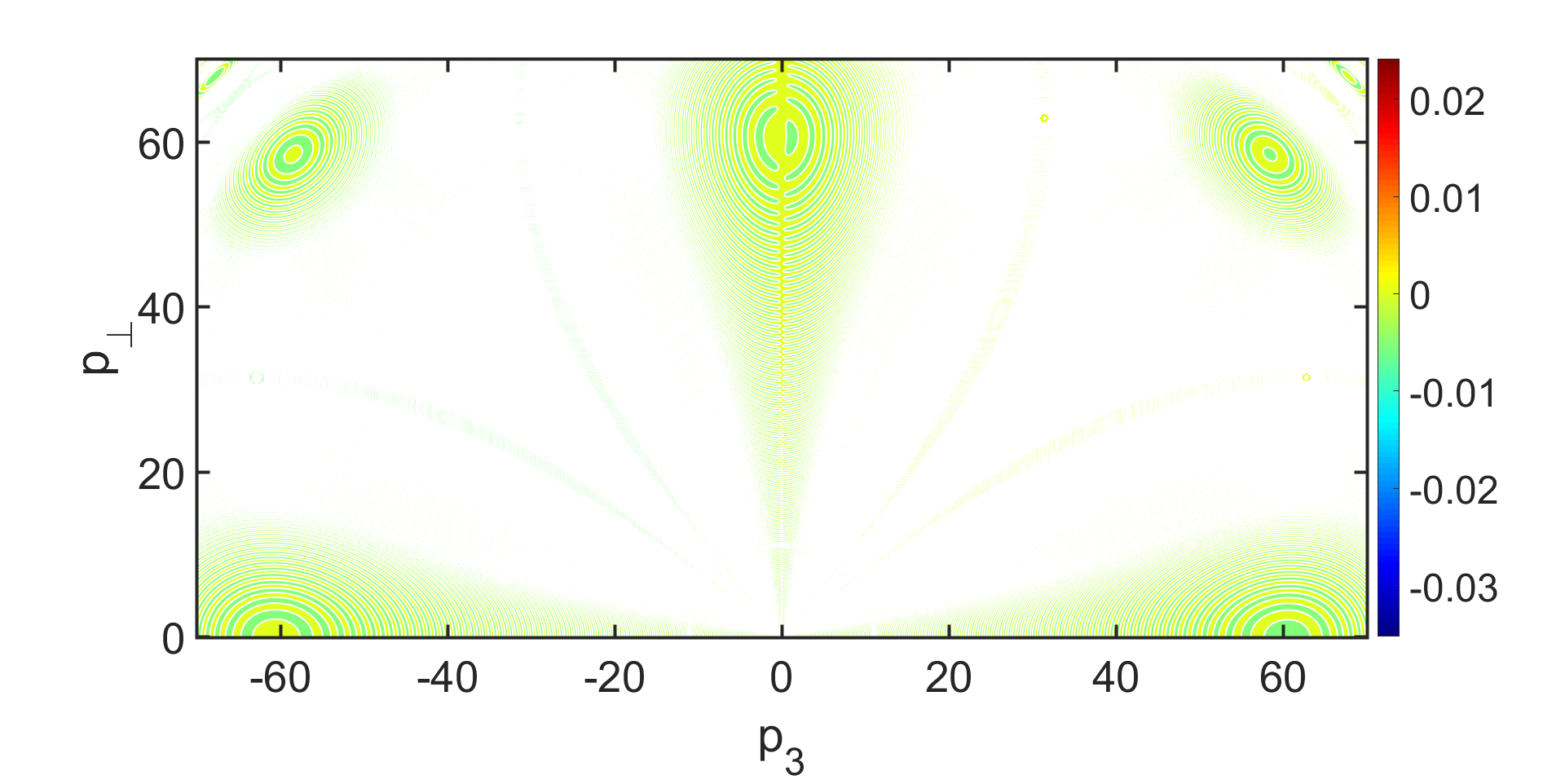}}\
  \subfloat[t=30.0]{
  \includegraphics[width=0.45\linewidth]{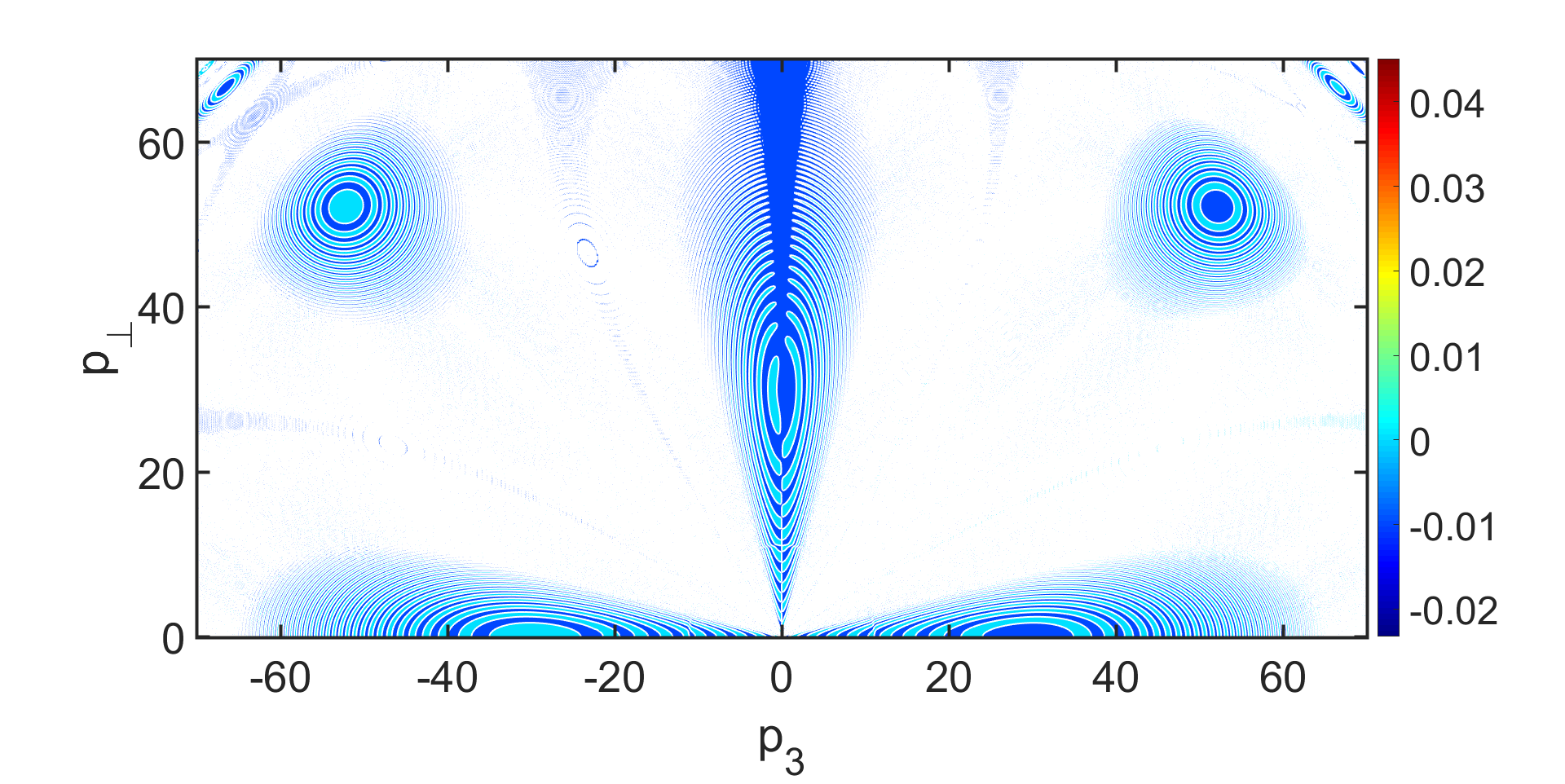}}\\
  \caption{(color on line) $b_{\perp}^{a}(p_{\perp},p_{3})$ at different times, where both $p_{\perp}$ and $p_{3}$ are in unit of $\frac{|E_{0}|}{\sqrt{E_{0}}}$}
  \label{fig7}
\end{figure}

\begin{figure}[H]
  \centering
  \subfloat[t=0.1]{
  \includegraphics[width=0.45\linewidth]{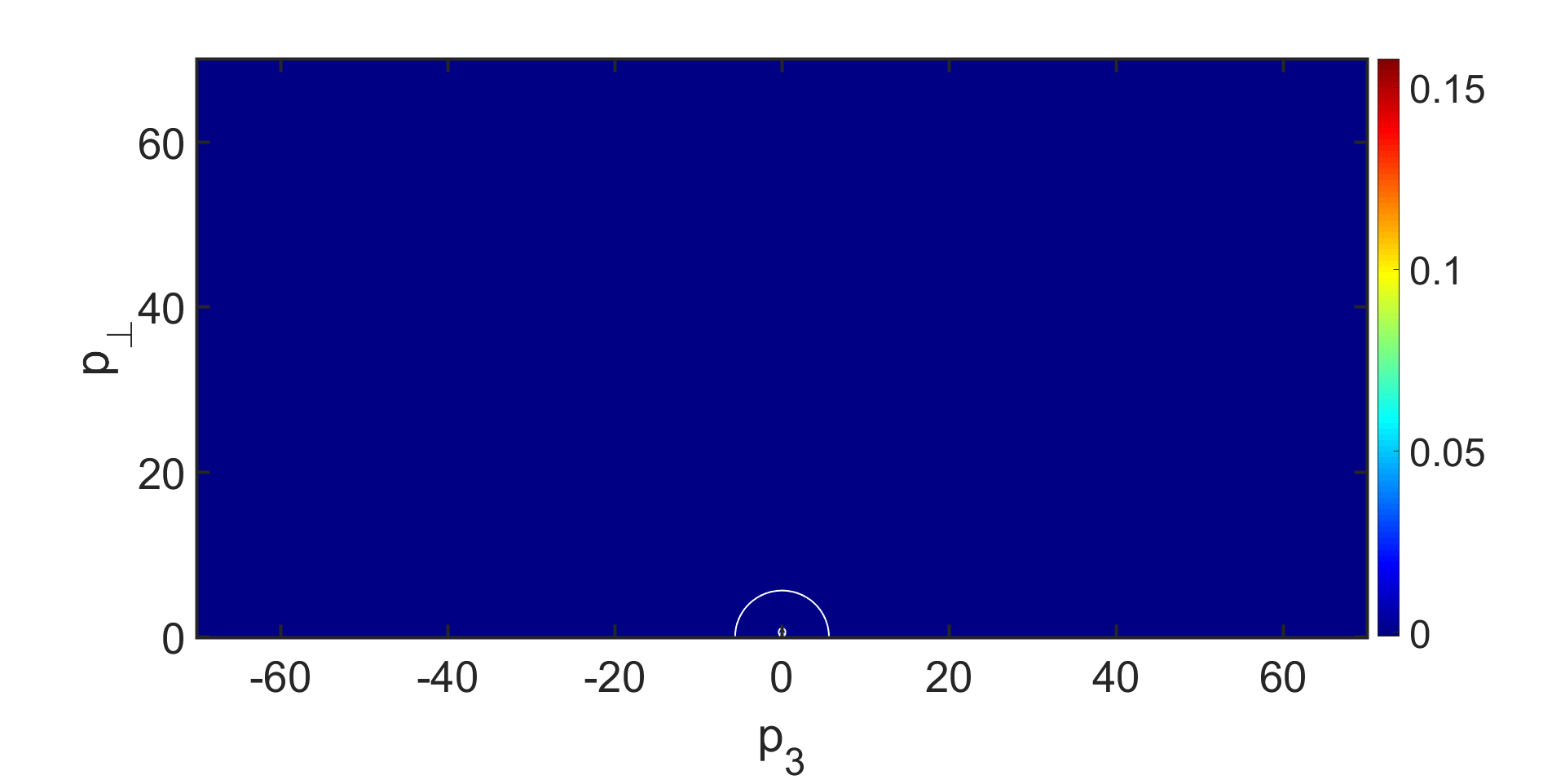}}\
  \subfloat[t=1.0]{
  \includegraphics[width=0.45\linewidth]{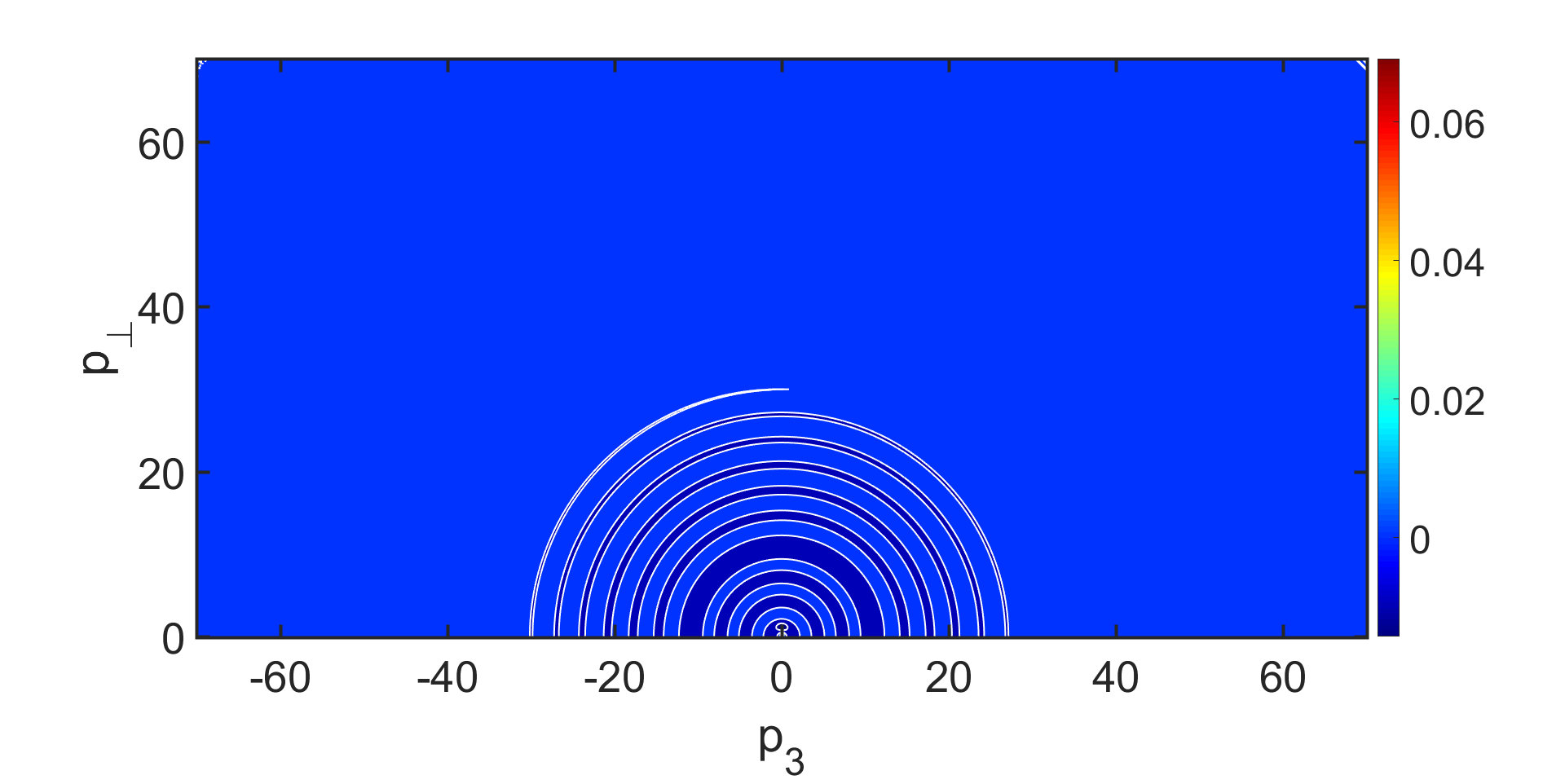}}\\
  \subfloat[t=5.0]{
  \includegraphics[width=0.45\linewidth]{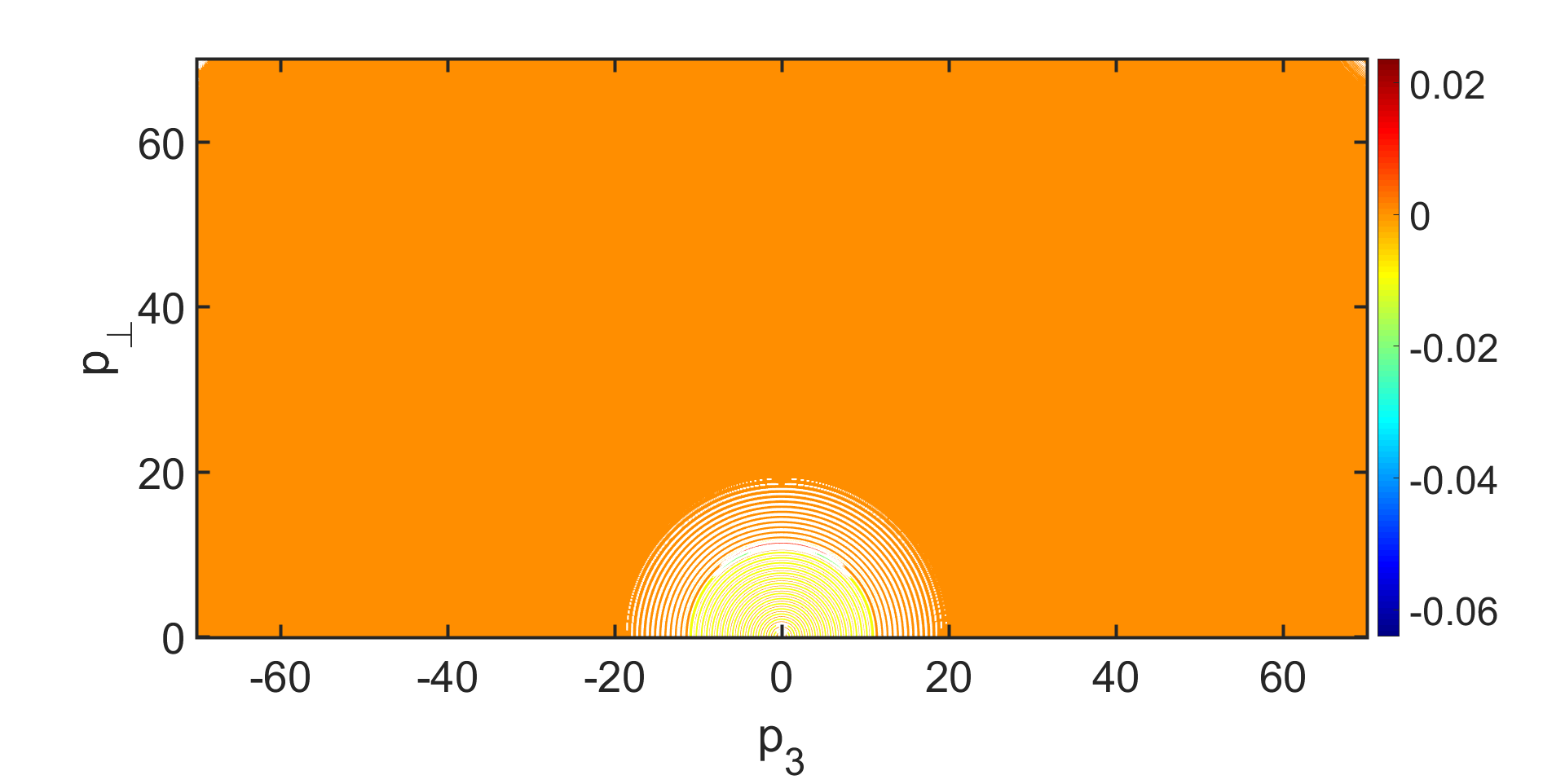}}\
  \subfloat[t=10.0]{
  \includegraphics[width=0.45\linewidth]{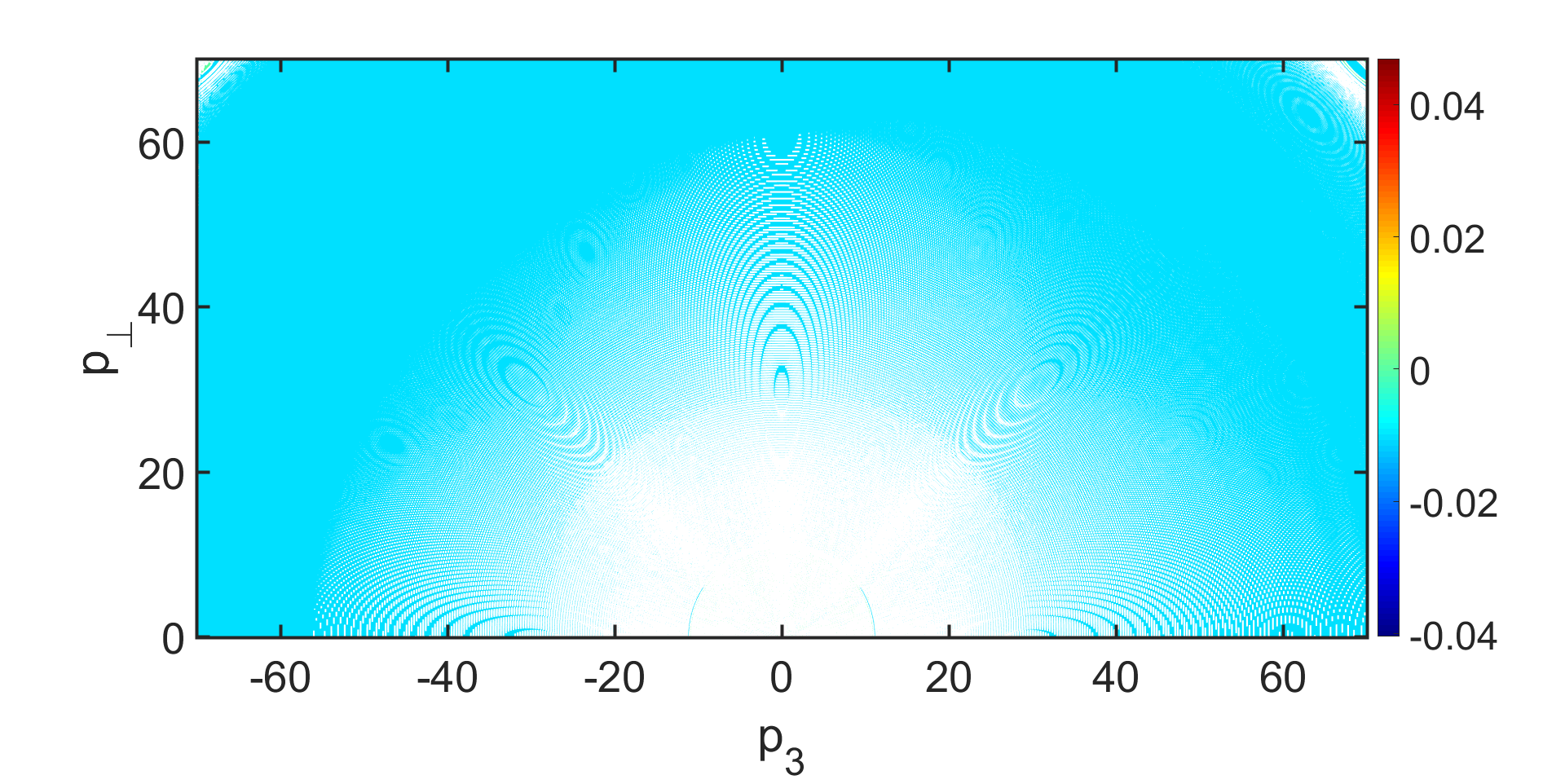}}\\
  \subfloat[t=15.0]{
  \includegraphics[width=0.45\linewidth]{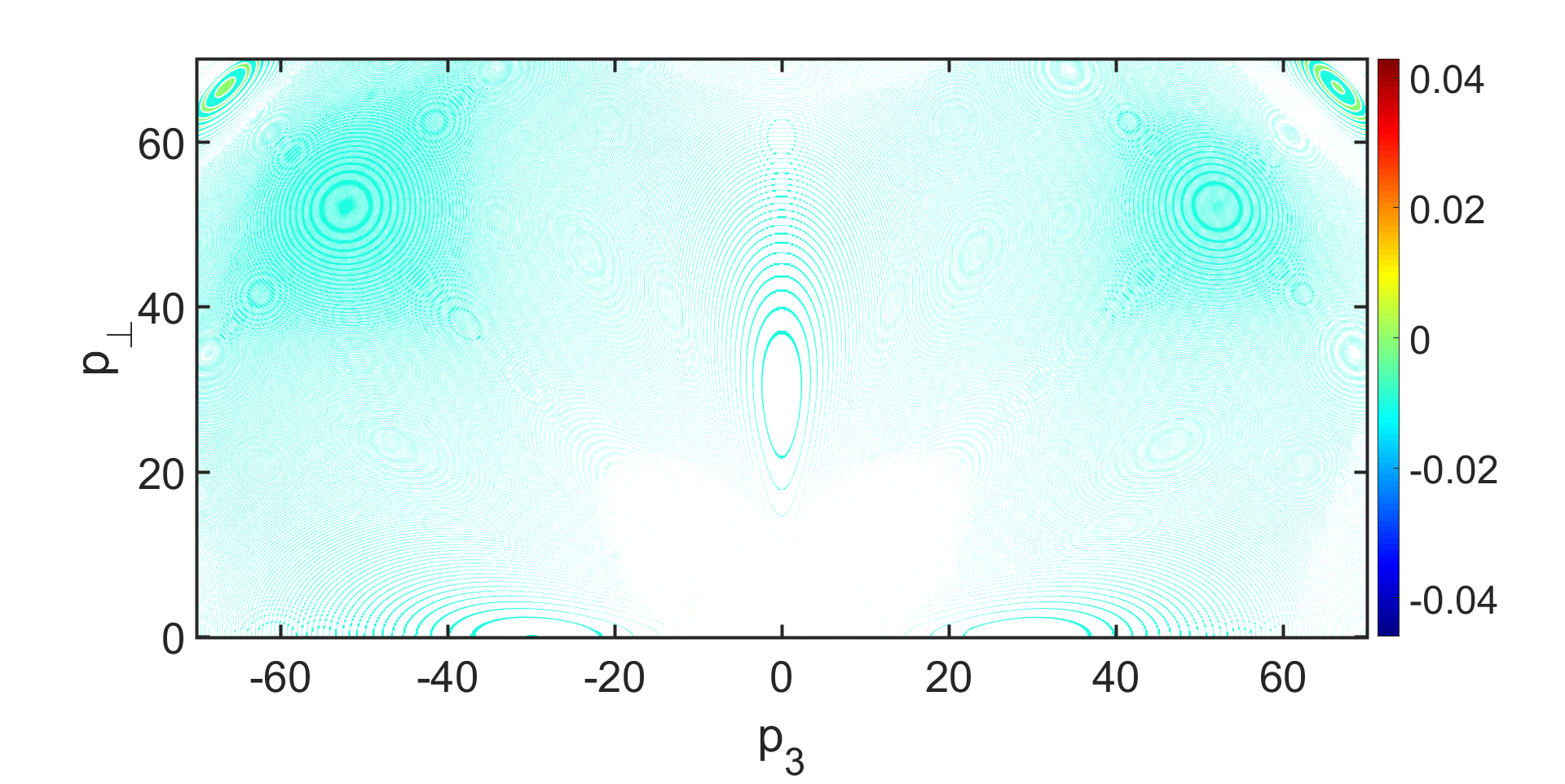}}\
  \subfloat[t=20.0]{
  \includegraphics[width=0.45\linewidth]{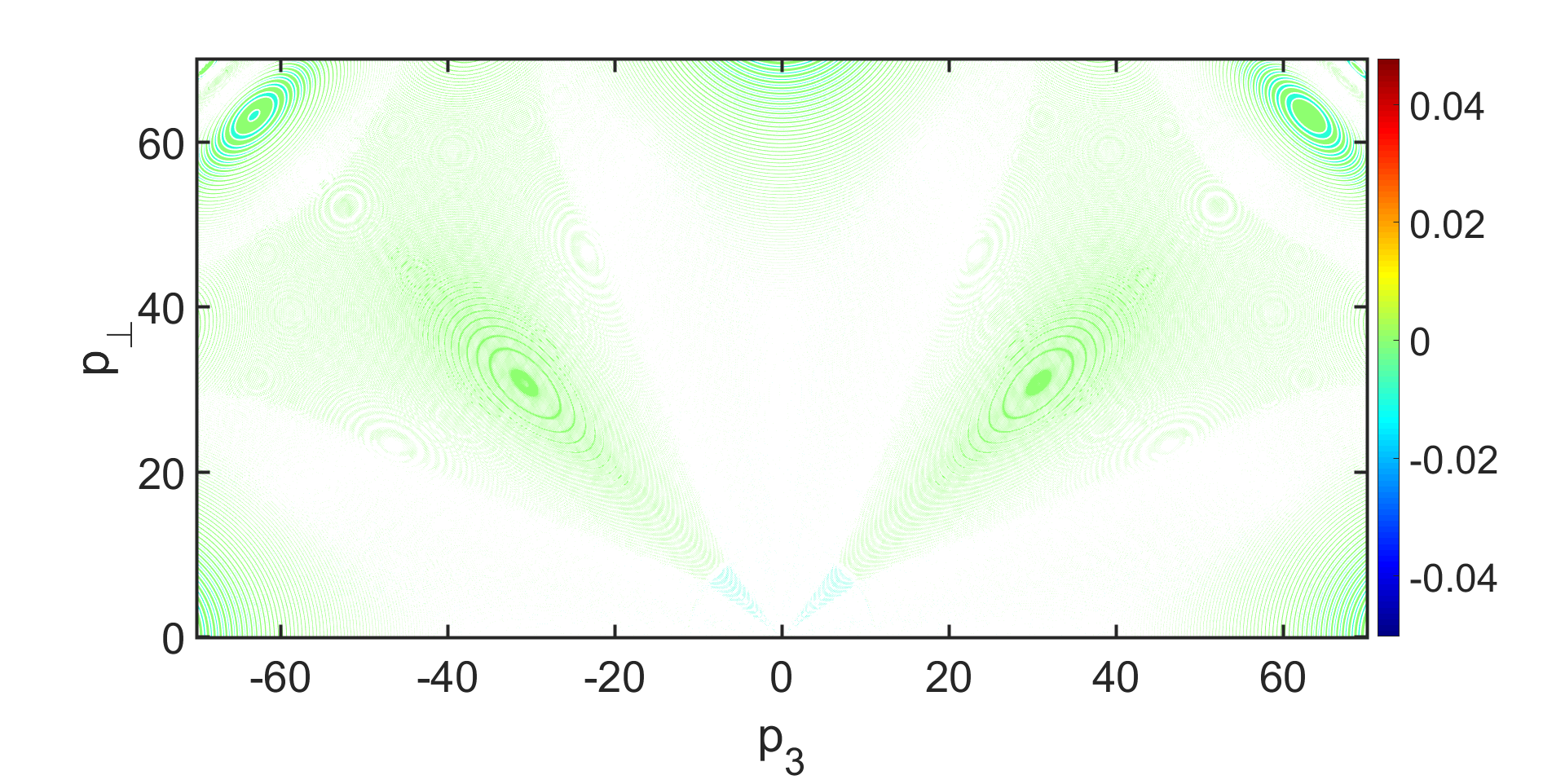}}\\
  \subfloat[t=25.0]{
  \includegraphics[width=0.45\linewidth]{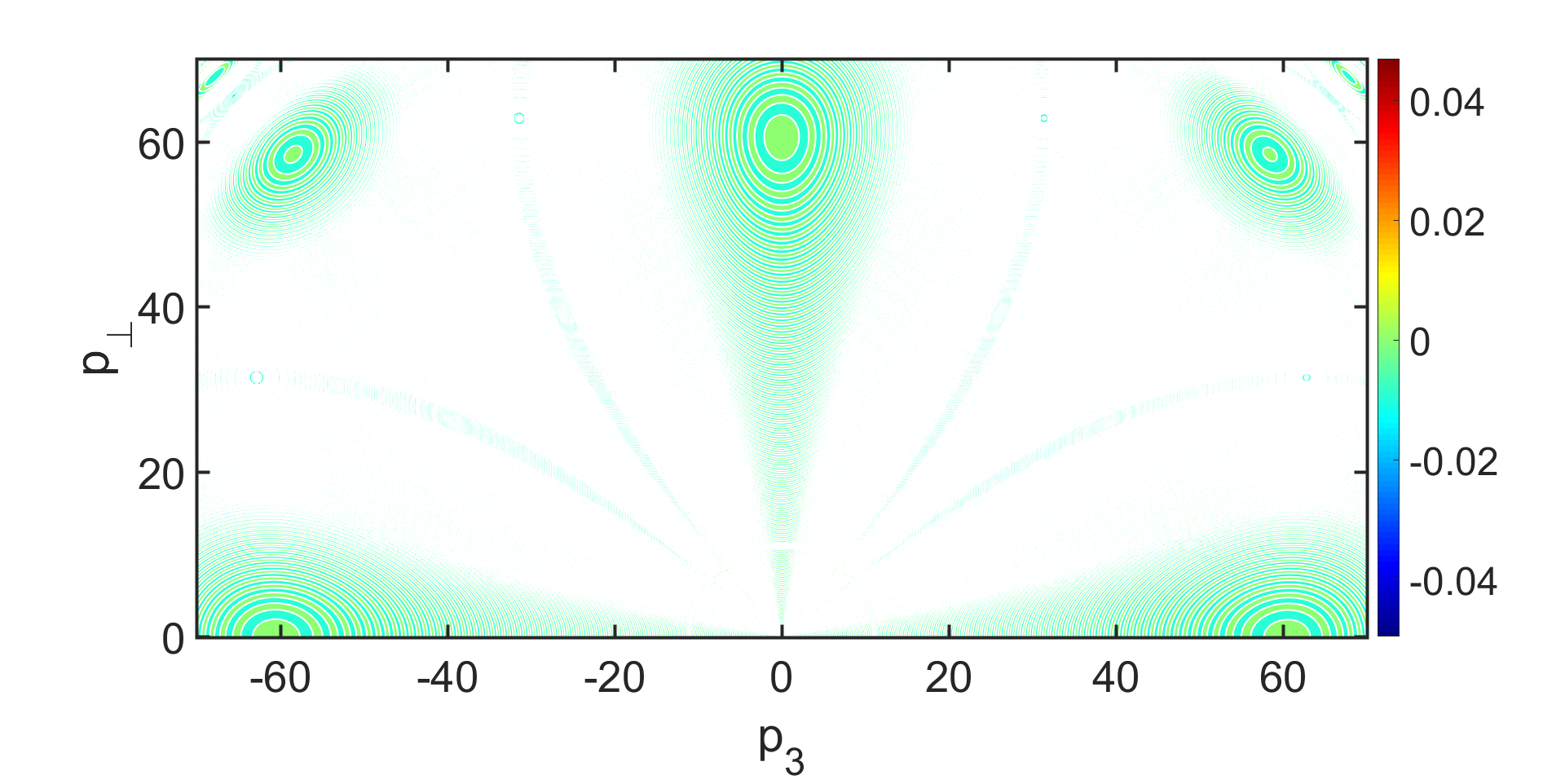}}\
  \subfloat[t=30.0]{
  \includegraphics[width=0.45\linewidth]{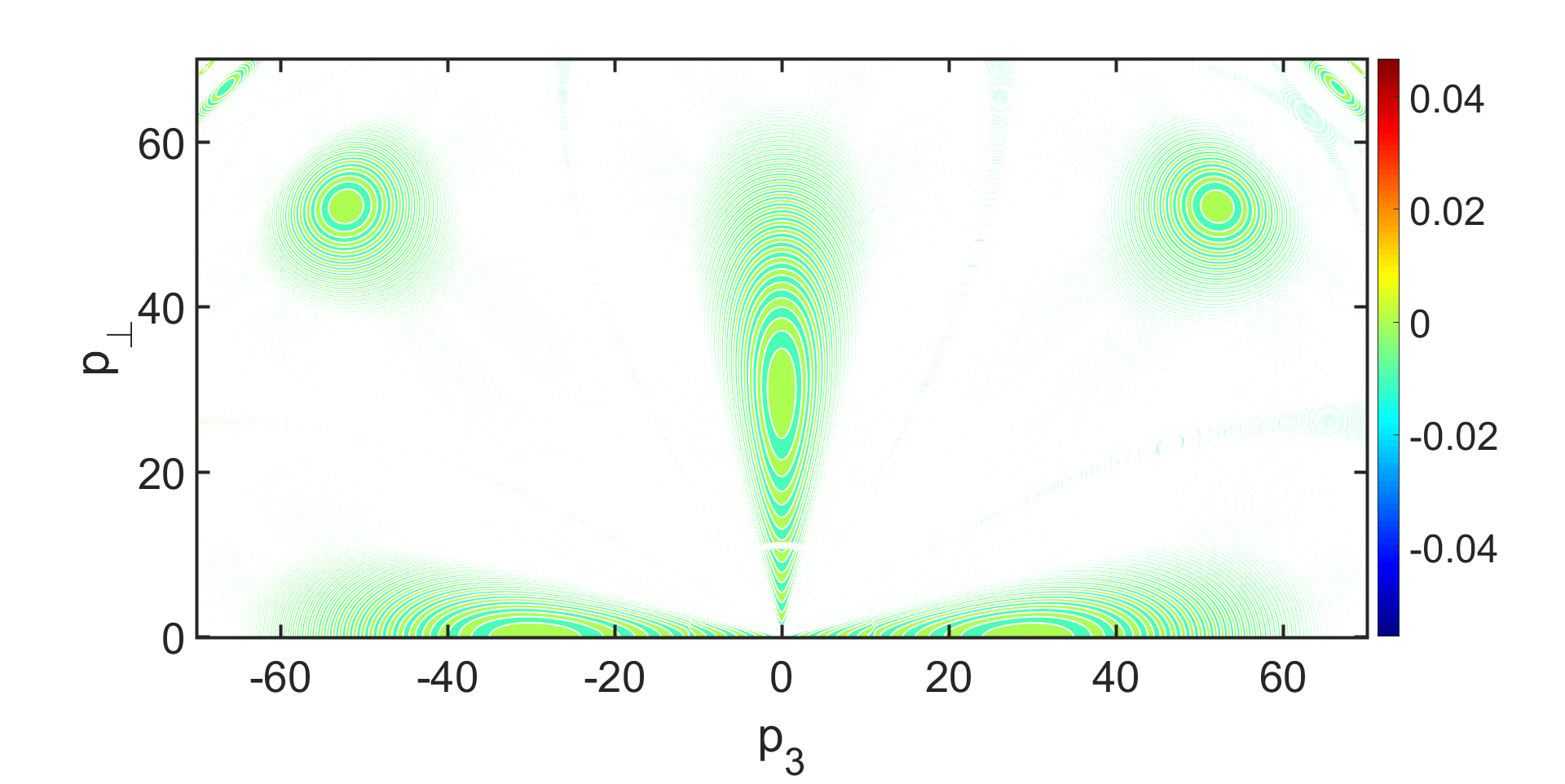}}\\
  \caption{(color on line) $b_{3}^{a}(p_{\perp},p_{3})$ at different times, where both $p_{\perp}$ and $p_{3}$ are in unit of $\frac{|E_{0}|}{\sqrt{E_{0}}}$}
  \label{fig8}
\end{figure}
To have an elementary understanding on the phenomenon, we try to analyze the components of the Wigner function, and plot $b_{\perp}^{s}(p_{\perp},p_{3})$, $b_{3}^{s}(p_{\perp},p_{3})$, $b_{\perp}^{a}$,$(p_{\perp},p_{3})$ and $b_{3}^{a}(p_{\perp},p_{3})$ in Fig.\ref{fig5}-Fig.\ref{fig8}.

Since the phenomenological distribution function relates to $b_{\perp}^{s}(p_{\perp},p_{3})$ and $b_{3}^{s}(p_{\perp},p_{3})$ directly from Eq.(\ref{eq30}), we plot singlet components at different time in Fig.\ref{fig5} and Fig.\ref{fig6}. It is obviously to see that $b_{\perp}^{s}(p_{\perp},p_{3})$ is symmetric along $p_{3}=0$ while $b_{3}^{s}(p_{\perp},p_{3})$ is  antisymmetric along $p_{3}=0$, which agrees with Ref.\cite{Skokov4}. The singlet components keep the same form at different time points. The multiplet components affect the current and field directly. Thus, they affect the quark production indirectly. In Fig.\ref{fig7} and Fig.\ref{fig8}, we plot $b_{\perp}^{a}(p_{\perp},p_{3})$ and $b_{3}^{a}(p_{\perp},p_{3})$. It can be seen that $b_{\perp}^{a}(p_{\perp},p_{3})$ is antisymmetric and $b_{3}^{a}(p_{\perp},p_{3})$ is symmetric along $p_{3}=0$, which act in the opposite way of the singlet components. As the time goes on,the interference-like structure appears, and spreads. At last, the components reach a steady state. The tendency of the variation mainly lie in three special directions, read $p_{3}/p_{\perp}\approx 0$, $p_{3}/p_{\perp}\approx \pm1$, $p_{3}/p_{\perp}\approx \infty$. Both the singlet and multiplet components couple with each other in the differential system, and take effect.

From Eq.(\ref{eq14})-Eq.(\ref{eq19}), when considering the back reaction, the field $E^{a}$ is related to the multiplet components $b_{\perp}^{a}$ and $b_{3}^{a}$, and the dependence of the singlet components on the multiplet components is changed, i.e. for two functions $G(b_{\perp}^{a},b_{3}^{a})$ and $H(b_{\perp}^{a},b_{3}^{a})$, have the relation $\frac{\partial G(b_{\perp}^{a},b_{3}^{a})}{\partial p_{3}} =E^{a}(b_{\perp}^{a},b_{3}^{a})\frac{\partial b^{a}_{\perp}}{\partial p_{3}}$ and  $\frac{\partial H(b_{\perp}^{a},b_{3}^{a})}{\partial p_{3}} =E^{a}(b_{\perp}^{a},b_{3}^{a})\frac{\partial b^{a}_{3}}{\partial p_{3}}$, then $G(b_{\perp}^{a},b_{3}^{a})$ and $H(b_{\perp}^{a},b_{3}^{a})$ have the properties as without considering the back reaction. In our opinion, the momentum gap and the confinement phenomenon are the back reaction effects in a coupled differential system.

\section{Conclusion}
In this work, we studied the massless quark production with considering the back reaction in $SU(2)$ gauge. We solved the kinetic equations of Wigner function numerically. When considering the back reaction, both the field and current are damping. Even though, the coupling act differently in nonperturbative models and perturbative models. The varying tendency of both field and current is reasonable. Comparing with the Bjorken expanding field, the yield of quark production is higher. In the distribution function, a momentum gap is existed, and the massless quark distribution forms a confinement phenomenon. Which, in our opinion, are caused by the back reaction. For these problems, the nonvanishing parts of the Wigner function, vector components, are qualitatively analyzed. The symmetrical and anti-symmetrical properties of the components are conserved in comparison with the Bjorken expanding field. The singlet components affect the distribution function directly while the multiplet components act indirectly. In return, the produce fermions form a feedback on the field. Our calculation showed the possibility to study a model of combining both nonperturbative and perturbative models. But to have a full understanding of the problem, the cross section should be carefully studied. Besides that, both the momentum gap and the confinement phenomenon also should be quantitatively investigated.

\begin{acknowledgments}
This work was supported by the National Natural Science Foundation of China (NSFC) under Grant No. 11475026. The authors would like to thank Dr. Marco Ruggieri for his important discussion. The computation was carried out at the HSCC of the Beijing Normal University.
\end{acknowledgments}

\end{document}